\documentclass[11pt]{article}
\usepackage{fullpage,amsthm}
\usepackage{amsmath,amsfonts,amssymb} 
\usepackage{feynmp,psfrag,graphicx}

\input xy
\xyoption{all}

\newtheorem{thm}{Theorem}
\newtheorem{corl}[thm]{Corollary}
\newtheorem{lma}[thm]{Lemma} 
\newtheorem{prop}[thm]{Proposition}
\newtheorem{defn}[thm]{Definition}
\newtheorem{ex}[thm]{Example}
\newtheorem{rem}[thm]{Remark}

\def\Aut{\mathrm{Aut}}
\def\bar{\overline}

\def\C{\mathbb{C}}

\def\Diff{\bar{\textup{Diff}}}

\def\F{\mathcal{F}}
\def\g{\mathfrak{g}}

\def\gh{\textup{gh}}

\def\Hom{\mathrm{Hom}}
\def\half{\tfrac{1}{2}}

\def\isom{\simeq}

\def\Loc{\textup{Loc}}

\def\N{\mathbb{N}} 

\def\op{\textup{op}}
\def\P{\mathcal{P}}
\def\R{\mathbb{R}}

\def\res{\mathrm{res}}

\def\rank{\textup{rank}}
\def\rk{\textup{rk}}

\def\Sym{\mathrm{Sym}}

\def\tr{\mathrm{tr~}}
\def\tilde{\widetilde}
\def\tot{\textup{tot}}
\def\U{\mathcal{U}}
\def\X{\mathfrak{X}}

\def\val{\mathrm{val}}
\def\Z{\mathbb{Z}}

\def\gho{
{~
      \begin{fmfgraph}(8,6)
	\fmfpen{.1mm}
	\fmfset{dot_len}{.5mm}
        \fmfforce{(0w,.15h)}{l}
        \fmfforce{(1w,.15h)}{r}
        \fmf{dots}{l,r}
    \end{fmfgraph}
}}

\def\glu{
{~
      \begin{fmfchar}(8,6)
	\fmfpen{.1mm}
	\fmfset{curly_len}{.5mm}
	\fmfleft{l}
	\fmfright{r}
	\fmf{gluon}{l,r}
      \end{fmfchar}
}}

\def\ghoglu{
{~
      \begin{fmfchar}(10,8)
	\fmfpen{.1mm}
	\fmfset{curly_len}{.5mm}
	\fmfset{dot_len}{.5mm}
	\fmfleft{l}
	\fmfright{r1,r2}
	\fmf{gluon}{l,v}
	\fmf{dots}{r1,v}
	\fmf{dots}{v,r2}
      \end{fmfchar}
}}

\def\gluc{
{~
      \begin{fmfchar}(10,8)
	\fmfpen{.1mm}
	\fmfset{curly_len}{.5mm}
	\fmfleft{l}
	\fmfright{r1,r2}
	\fmf{gluon}{l,v}
	\fmf{gluon}{v,r1}
	\fmf{gluon}{v,r2}
      \end{fmfchar}
}}

\def\gluq{
{~
      \begin{fmfchar}(8,6)
	\fmfpen{.1mm}
	\fmfset{curly_len}{.5mm}
	\fmfleft{l1,l2}
	\fmfright{r1,r2}
	\fmf{gluon}{l1,v}
	\fmf{gluon}{l2,v}
	\fmf{gluon}{r1,v}
	\fmf{gluon}{v,r2}
      \end{fmfchar}
}}

\def\gluBRST{
{~
      \begin{fmfchar}(10,8)
	\fmfpen{.1mm}
	\fmfset{curly_len}{.5mm}
	\fmfleft{l}
	\fmfright{r1,r2}
	\fmf{dashes}{l,v}
	\fmf{dots}{v,r1}
	\fmf{gluon}{v,r2}
      \end{fmfchar}
}}

\def\ghoBRST{
{~
      \begin{fmfchar}(10,8)
	\fmfpen{.1mm}
	\fmfset{curly_len}{.5mm}
	\fmfleft{l}
	\fmfright{r1,r2}
	\fmf{dashes}{l,v}
	\fmf{dots}{v,r1}
	\fmf{dots}{v,r2}
      \end{fmfchar}
}}

\title{The structure of renormalization Hopf algebras for gauge theories I: Representing Feynman graphs on BV-algebras}
\author{Walter D. van Suijlekom\\[3mm]
Institute for Mathematics, Astrophysics and Particle Physics\\
Faculty of Science, Radboud Universiteit Nijmegen\\
Toernooiveld 1, 6525 ED Nijmegen, The Netherlands\\[3mm]
\texttt{waltervs@math.ru.nl}}

\date{February 9, 2009}

\begin{document}
\begin{fmffile}{graphs-nola}

\fmfset{wiggly_len}{5pt} 
\fmfset{wiggly_slope}{70} 
\fmfset{curly_len}{4pt} 
\fmfset{curly_len}{1.4mm}

\fmfset{dot_len}{1mm}

\maketitle

\begin{abstract}
We study the structure of renormalization Hopf algebras of gauge theories. We identify certain Hopf subalgebras in them, whose character groups are semidirect products of invertible formal power series with formal diffeomorphisms. This can be understood physically as wave function renormalization and renormalization of the coupling constants, respectively. 
After taking into account the Slavnov--Taylor identities for the couplings as generators of a Hopf ideal, we find Hopf subalgebras in the corresponding quotient as well.

In the second part of the paper, we explain the origin of these Hopf ideals by considering a coaction of the renormalization Hopf algebras on the Batalin-Vilkovisky (BV) algebras generated by the fields and couplings constants. 
The so-called classical master equation satisfied by the action in the BV-algebra implies the existence of the above Hopf ideals in the renormalization Hopf algebra. Finally, we exemplify our construction by applying it to Yang--Mills gauge theory. 
\end{abstract}

\section{Introduction}
\label{sect:intro}
The mathematical formulation of quantum gauge theories forms one of the great challenges in mathematical physics. Recently, the perturbative structure of quantum Yang--Mills gauge theories has been more and more understood. On the one hand, many rigorous results can be obtained \cite{BBH94,BBH95} using cohomological arguments within the context of the BRST-formalism \cite{BRS74,BRS75,BRS76,Tyu75}. On the other hand, renormalization of perturbative quantum field theories has been carefully structured using Hopf algebras \cite{Kre98,CK99,CK00}. The presence of a gauge symmetry induces a rich additional structure on these Hopf algebras, as has been explored in \cite{Kre05,KY06,BKUY08} and in the author's own work \cite{Sui07c,Sui07}. All of this work is based on the algebraic transparency of BPHZ-renormalization, with the Hopf algebra reflecting the recursive nature of this procedure. 

Nevertheless, there are two objections to this approach to perturbative quantum field theories. Firstly, it is defined in momentum space and one is thus restricted to quantum field theories on flat spacetime and, secondly, it is defined as a graph-by-graph procedure and not in terms of the -- more physical -- full Green's functions. In this paper, we will address the second point and try to elucidate the Hopf algebraic structure on the level of Green's functions in gauge theories. 
The first point has been addressed in the series of papers \cite{BruF00,DF00,HW02} and references therein, in the context of algebraic quantum field theories. The case of Yang--Mills theories was considered by Hollands in \cite{Hol07}. Interesting to note is that the Epstein--Glaser renormalization involved in this approach has an underlying Hopf algebraic structure as well \cite{Pin00} (see also \cite{BK04} for rooted trees instead of Feynman graphs).

The mainstream physics literature has taken a slightly different road to the perturbative approach to quantum gauge theories by putting functional integration techniques at its heart. Although the formal path integral manipulations provide a powerful technique, it is hard to overestimate the importance of a transparent algebraic description of perturbative quantum gauge theories. 

In this paper, we adopt the philosophy to put formal (perturbative) expansions in Feynman graphs as a starting point and try to (rigorously) derive results from that, avoiding functional techniques completely. For instance, the process of (BPHZ)-renormalization is captured by means of the Connes--Kreimer Hopf algebra. As we will explain in Section \ref{sect:prel} below, this extends naturally to gauge theories. The Slavnov--Taylor identities for the couplings -- the reminiscences of the gauge symmetry -- are shown to be compatible with renormalization by establishing them as generators of a Hopf ideal in this Hopf algebra (Section \ref{sect:setup}). In the corresponding quotient Hopf algebra, we find certain Hopf subalgebras and we will show that their character group is a subgroup of the semidirect product of (invertible) formal power series with a formal diffeomorphism group. It should be noted that the existence of Hopf subalgebras has already been studied in the context of Dyson--Schwinger equations in \cite{BK05,Kre05} and in \cite{Foi07} for planar rooted trees. Also, our work reflects the semi-direct product considered in \cite{GKM01} for a scalar field theory, and \cite{BF01,BFK06} where the above groups of formal series and diffeomorphisms (and their noncommutative analogues) appears in the study of renormalization of quantum electrodynamics.

Although the existence of the Hopf ideals can be established rigorously by a combinatorial proof, it is crucial to have a more conceptual understanding of their origin. This is what we do in the second part of the paper. Namely, we connect the renormalization Hopf algebras of gauge theories to the BV-algebras generated by the fields and coupling constants by making the latter a comodule BV-algebra over the Hopf algebras  (Section \ref{sect:coaction}). The induced action of the character group can be understood as wave function renormalization (invertible formal power series in $k$ variables) and coupling constant renormalization (formal diffeomorphisms of $\C^k$). The origin of the Hopf ideals is clarified by identifying an ideal in the BV-algebra generated by the so-called master equation satisfied by the action. This ideal induces a Hopf ideal in the renormalization Hopf algebra which in the case of simple gauge theories coincides with the ideal generated by the Slavnov--Taylor identities for the couplings. On the level of the character group, the $k$ parameters reduce to the subgroup in one parameter. This reflects the presence of a `fundamental coupling constant' in such theories, in terms of which all other coupling constants can be expressed. We conclude this section by discussing the renormalization group and the beta-function in this context. In Section \ref{sect:ym} we exemplify our construction by working out explicitly the case of Yang--Mills gauge theory, with a simple Lie group. 

The study of the effective action of gauge theories will be postponed to a second paper. In particular, we will study the Zinn--Justin equation from the Hopf-algebraic point of view as well as its relation with the (classical) master equation (cf. Eq. \eqref{eq:master} below) satisfied by the action.

\section{Preliminaries on Hopf algebras}
\label{sect:prel}
Let us briefly recall the role played by Hopf algebras as coordinate rings on (affine) groups, while referring to \cite{Wat79} for a complete treatment.  

We will consider commutative algebras $H$ (over $\C$) for which the set of characters $G:=\Hom_\C(H,\C)$ is actually a group. The group structure on $G$ induces the structure of a {\it Hopf algebra} on $H$, that is, a counit, $\eta: H \to \C$, a coproduct $\Delta: H \to H \otimes H$ and an antipode $S: H \to H$. These are all algebra maps and are supposed to satisfy certain compatibility condition which we do not list here. 

Representations of $G$ correspond one-to-one to {\it corepresentations} of $H$. In fact, if $V$ is a $G$-module, then it is also a {\it comodule} over $H$, that is, there exists a map (called {\it coaction}) $\rho : V \to V \otimes H$ such that $g v = (1 \otimes g) \rho(v)$. If $V$ has additional structure, it is natural to require the coaction to respect this structure.

We will further restrict our study to {\it connected graded Hopf algebras} for which there is a grading $H=\oplus_{n \in \N} H^n$ that is respected by the product and the coproduct:
\begin{align*} 
H^k H^l \subset H^{k+l}; \qquad
\Delta(H^n) = \sum_{k=0}^n H^k \otimes H^{n-k}.
\end{align*}
and such that $H^0 = \C 1$. Dually, graded Hopf algebras correspond to (pro)-unipotent groups. We illustrate the above definition somewhat elaborately by discussing in the next two subsections two examples that are relevant in what follows. 

\subsection{Hopf algebra of Feynman graphs}
We suppose that we have defined a (renormalizable) perturbative quantum field theory and specified the possible interactions between different types of fields. These fields are collected in a set $\Phi=\{ \phi_1, \ldots, \phi_{N'} \}$ whereas the different types of interactions -- represented by vertices -- constitute a set $R_V$. In the Lagrangian formalism, it is natural to associate to each vertex a local monomial in the fields (present in the Lagrangian); we will denote this map by $\iota : R_V \to \Loc(\Phi)$, where $\Loc(\Phi)$ is defined as the algebra of local polynomials in the fields $\phi_j$ (see Definition \ref{defn:localform} below).

Propagators, on the other hand, are indicated by edges and form a set $R_E$. Again, one assigns a monomial to each edge via $\iota : R_E \to \Loc(\Phi)$ but now $\iota(e)$ which is now of order 2 in the fields, involving precisely the field (and its conjugate in the case of fermions) that is propagating.

We will assume that there are BRST-source terms present in the theory, which means that for each field $\phi_i$ there is a corresponding source field $K_{\phi_i}$ in $\Phi$. In other words, the set of fields is of the form
$$
\Phi = \{ \phi_1, \cdots, \phi_N, K_{\phi_1}, \cdots, K_{\phi_N} \}
$$
This {\it even-dimensionality} is a manifestation of the structure on the fields of a Gerstenhaber algebra which we will explore later.

\begin{ex}
Quantum electrodynamics describes the interaction of charged particles such as electrons with photons, with corresponding fields $\psi$ and $A$. Their propagation is usually indicated by a straight and a wiggly line (for the electron and photon, respectively). There is only the interaction of an electron emitting a photon: this is indicated by a vertex of valence three; the mass term for the electron is indicated by a vertex of valence two. The dynamical and interactive character of the theory can be summarized by the following sets,\footnote{We specify the type of fields that are involved in the interaction by drawing a small neighborhood around the vertex instead of merely a dot.}
\begin{align*}
R_V =\{~ 
\parbox{20pt}{
    \begin{fmfgraph}(20,10)
      \fmfleft{l}
      \fmfright{r1,r2}
      \fmf{photon}{l,v}
      \fmf{plain}{r1,v}
      \fmf{plain}{v,r2}
      \fmfdot{v}
    \end{fmfgraph}}~,~
\parbox{20pt}{
  \begin{fmfgraph}(20,10)
      \fmfleft{l}
      \fmfright{r}
      \fmf{plain}{l,v}
      \fmf{plain}{v,r}
      \fmfdot{v}
    \end{fmfgraph}}
~\};
\qquad
R_E = \{~
\parbox{20pt}{
    \begin{fmfgraph}(20,10)
      \fmfleft{l}
      \fmfright{r}
      \fmf{plain}{l,r}
    \end{fmfgraph}}~,~
\parbox{20pt}{
    \begin{fmfgraph}(20,10)
      \fmfleft{l}
      \fmfright{r}
      \fmf{photon}{l,r}
    \end{fmfgraph}}~\}.
\end{align*}
The corresponding monomials in $\F([\Phi])$ are
\begin{gather*}
\iota (\parbox{20pt}{\begin{fmfgraph}(20,10)
      \fmfleft{l}
      \fmfright{r1,r2}
      \fmf{photon}{l,v}
      \fmf{plain}{r1,v}
      \fmf{plain}{v,r2}
      \fmfdot{v}
    \end{fmfgraph}}) = - e \bar\psi \gamma\circ A \psi, \qquad 
\iota(\parbox{20pt}{ \begin{fmfgraph}(20,10)
      \fmfleft{l}
      \fmfright{r}
      \fmf{plain}{l,v}
      \fmf{plain}{v,r}
      \fmfdot{v}
    \end{fmfgraph}}) = -m \bar\psi \psi, \\
\iota ( \parbox{20pt}{
    \begin{fmfgraph}(20,10)
      \fmfleft{l}
      \fmfright{r}
      \fmf{plain}{l,r}
    \end{fmfgraph}} ) =i \bar\psi \gamma \circ d \psi
,\qquad
\iota (\parbox{20pt}{
    \begin{fmfgraph}(20,10)
      \fmfleft{l}
      \fmfright{r}
      \fmf{photon}{l,r}
    \end{fmfgraph}}) = -d A * dA.
\end{gather*}
with $e$ and $m$ the electric charge and mass of the electron, respectively.
\end{ex}
\begin{ex}
Quantum chromodynamics describes the strong interaction between quarks and gluons, described by the fields $\psi$ and $A$, respectively (see Section \ref{sect:ym} below for more details). These are indicated by straight and wiggly lines. In addition, associated to the non-abelian gauge symmetry (with symmetry group $SU(3)$) there is the so-called ghost field $\omega$, indicated by dotted lines, as well as the BRST-sources $K_\psi, K_A$ and $K_\omega$. Between the fields there are four interactions, three BRST-source terms, and a mass term for the quark. This leads to the following sets of vertices and edges,
\begin{align*}
R_V &= \left\{ 
\raisebox{-7.5pt}{
\parbox{20pt}{
    \begin{fmfchar}(20,15)
      \fmfleft{l}
      \fmfright{r1,r2}
      \fmf{gluon}{l,v}
      \fmf{plain}{r1,v}
      \fmf{plain}{v,r2}
      \fmfdot{v}
    \end{fmfchar}
  }}
,
\raisebox{-7.5pt}{
\parbox{20pt}{
  \begin{fmfgraph}(20,15)
      \fmfleft{l}
      \fmfright{r1,r2}
      \fmf{gluon}{l,v}
      \fmf{dots}{r1,v}
      \fmf{dots}{v,r2}
      \fmfdot{v}
  \end{fmfgraph}
}}
,
\raisebox{-7.5pt}{
\parbox{20pt}{
  \begin{fmfgraph}(20,15)
    \fmfleft{l}
      \fmfright{r1,r2}
      \fmf{gluon}{l,v}
      \fmf{gluon}{r1,v}
      \fmf{gluon}{v,r2}
      \fmfdot{v}
  \end{fmfgraph}
}}
,
\raisebox{-7.5pt}{
\parbox{20pt}{
  \begin{fmfgraph}(20,15)
    \fmfleft{l1,l2}
      \fmfright{r1,r2}
      \fmf{gluon}{l1,v}
      \fmf{gluon}{l2,v}
      \fmf{gluon}{r1,v}
      \fmf{gluon}{v,r2}
      \fmfdot{v}
  \end{fmfgraph}
}}~,~
\raisebox{-7.5pt}{
\parbox{20pt}{
    \begin{fmfchar}(20,15)
      \fmfleft{l}
      \fmfright{r1,r2}
      \fmf{dashes}{l,v}
      \fmf{dots}{r1,v}
      \fmf{plain}{v,r2}
      \fmfdot{v}
    \end{fmfchar}
}}~,~
\raisebox{-7.5pt}{
\parbox{20pt}{
    \begin{fmfchar}(20,15)
      \fmfleft{l}
      \fmfright{r1,r2}
      \fmf{dashes}{l,v}
      \fmf{dots}{r1,v}
      \fmf{gluon}{v,r2}
      \fmfdot{v}
    \end{fmfchar}
}}~,~
\raisebox{-7.5pt}{
\parbox{20pt}{
    \begin{fmfchar}(20,15)
      \fmfleft{l}
      \fmfright{r1,r2}
      \fmf{dashes}{l,v}
      \fmf{dots}{r1,v}
      \fmf{dots}{v,r2}
      \fmfdot{v}
    \end{fmfchar}
}}~,~
\parbox{15pt}{
  \begin{fmfchar}(15,5)
      \fmfleft{l}
      \fmfright{r}
      \fmf{plain}{l,v}
      \fmf{plain}{v,r}
      \fmfdot{v}
    \end{fmfchar}}
\right\}
\intertext{with the dashed lines representing the BRST-source terms, and}
R_E &= \left\{ 
\raisebox{-7.5pt}{
\parbox{20pt}{
  \begin{fmfgraph}(20,10)
      \fmfleft{l}
      \fmflabel{}{l}
      \fmfright{r}
      \fmf{plain}{l,r}
  \end{fmfgraph}
}}
~,~
\raisebox{-7.5pt}{
\parbox{20pt}{
  \begin{fmfgraph}(20,10)
      \fmfleft{l}
      \fmflabel{}{l}
      \fmfright{r}
      \fmf{dots}{l,r}
  \end{fmfgraph}
}}
~,~
\raisebox{-7.5pt}{
\parbox{20pt}{
  \begin{fmfgraph}(20,10)
      \fmfleft{l}
      \fmflabel{}{l}
      \fmfright{r}
      \fmf{gluon}{l,r}
  \end{fmfgraph}
}}
\right\}.
\end{align*}
Note that the dashed edges do not appear in $R_E$, {i.e.} the source terms do not propagate and in the following will not appear as internal edges of a Feynman graph.
\end{ex}
\begin{rem}
\label{rem:psipsibar}
Although these examples motivate our construction, we stress that for what follows it is not necessary to specify the fields in the set $\Phi$ nor the vertices and edges in $R= R_V \cup R_E$ explicitly. 
The relevant structure is encoded by the map $\iota : R \to \Loc(\Phi)$. We note, however, that we make the following natural working assumptions:
\begin{enumerate}
\item Whenever a fermionic field, say $\psi$, interacts at a vertex $v \in R_V$ which does not involve a BRST-source, then $\iota(v)$ involves {\it both} $\psi$ and $\bar\psi$.
\item There is only one vertex for every BRST-source.
\item There are no valence two vertices involving two different fields (thus, still allowing mass terms).
\end{enumerate}
Physically, the last condition means that we require order two polynomials other than mass terms in the Lagrangian not to be radiatively corrected.
\end{rem}

A {\it Feynman graph} is a graph built from the types of vertices present in $R_V$ and the types of edges present in $R_E$. Naturally, we demand edges to be connected to vertices in a compatible way, respecting the type of vertex and edge. As opposed to the usual definition in graph theory, Feynman graphs have no external vertices. However, they do have {\it external lines} which come from vertices in $\Gamma$ for which some of the attached lines remain vacant ({i.e.} no edge in $R_E$ attached). Implicit in the construction is the fact that source terms only arise as external lines since they are not in $R_E$, justifying the name 'source term'.

If a Feynman graph $\Gamma$ has two external lines, both corresponding to the same field, we would like to distinguish between propagators and mass terms. In more mathematical terms, since we have vertices of valence two, we would like to indicate whether a graph with two external lines corresponds to such a vertex, or to an edge. A graph $\Gamma$ with two external lines is dressed by a bullet when it corresponds to a vertex, {i.e.} we write $\Gamma_\bullet$. The above correspondence between Feynman graphs and vertices/edges is given by the {\it residue} $\res(\Gamma)$. 
It is defined as the vertex or edge the graph corresponds to after collapsing all its internal points. 
For example, we have:
\begin{gather*}
\res\left( \parbox{40pt}{
\begin{fmfgraph*}(40,30)
      \fmfleft{l}
      \fmfright{r1,r2}
      \fmf{photon}{l,v}
      \fmf{plain}{v,v1,r1}
      \fmf{plain}{v,v2,r2}
      \fmffreeze
      \fmf{photon}{v1,loop,v2}
      \fmffreeze
      \fmfv{decor.shape=circle, decor.filled=0, decor.size=2thick}{loop}
\end{fmfgraph*}
}\right) = 
\parbox{20pt}{
\begin{fmfgraph*}(20,20)
      \fmfleft{l}
      \fmfright{r1,r2}
      \fmf{photon}{l,v}
      \fmf{plain}{v,r1}
      \fmf{plain}{v,r2}
\end{fmfgraph*}
}
\qquad \text{ and }\qquad
\res\left( \parbox{40pt}{
\begin{fmfgraph*}(40,30)
      \fmfleft{l}
      \fmfright{r}
      \fmf{plain}{l,v1,v2,v5,v6,r}
      \fmf{photon,left,tension=0}{v1,v5}
      \fmf{photon,right,tension=0}{v2,v6}
\end{fmfgraph*}
}\right) = 
\parbox{20pt}{
\begin{fmfgraph*}(20,20)
      \fmfleft{l}
      \fmfright{r}
      \fmf{plain}{l,r}
\end{fmfgraph*}}~,~
\quad \text{ but: }\qquad
\res\left( \parbox{40pt}{
\begin{fmfgraph*}(40,30)
      \fmfleft{l}
      \fmfright{r}
      \fmf{plain}{l,v1,v2,v5,v6,r}
      \fmf{photon,left,tension=0}{v1,v5}
      \fmf{photon,right,tension=0}{v2,v6}
\end{fmfgraph*}}{}_\bullet
\right) = 
\parbox{20pt}{
\begin{fmfgraph*}(20,20)
      \fmfleft{l}
      \fmfright{r}
      \fmf{plain}{l,v,r}
      \fmfdot{v}
\end{fmfgraph*}}
\end{gather*}

For the definition of the Hopf algebra of Feynman graphs, we restrict to {\it one-particle irreducible} (1PI) Feynman graphs. These are graphs that are not trees and cannot be disconnected by cutting a single internal edge. 
\begin{defn}[Connes--Kreimer \cite{CK99}]
The Hopf algebra of Feynman graphs is the free commutative algebra $H$ over $\C$ generated by all 1PI Feynman graphs with residue in $R= R_V \cup R_E$, with counit $\epsilon(\Gamma)=0$ unless $\Gamma=\emptyset$, in which case $\epsilon(\emptyset)=1$, coproduct,
\begin{align*}
\Delta (\Gamma) = \Gamma \otimes 1 + 1 \otimes \Gamma + \sum_{\gamma \subsetneq \Gamma} \gamma \otimes \Gamma/\gamma,
\end{align*}
where the sum is over disjoint unions of 1PI subgraphs with residue in $R$. The quotient $\Gamma/\gamma$ is defined to be the graph $\Gamma$ with the connected components of the subgraph contracted to the corresponding vertex/edge. If a connected component $\gamma'$ of $\gamma$ has two external lines, then there are possibly two contributions corresponding to the valence two vertex and the edge; the sum involves the two terms $\gamma'_\bullet \otimes \Gamma/(\gamma' \to \bullet)$ and $\gamma' \otimes \Gamma/\gamma'$.
The antipode is given recursively by,
\begin{equation*}
S(\Gamma) = - \Gamma - \sum_{\gamma \subsetneq \Gamma} S(\gamma) \Gamma/\gamma.
\end{equation*}
\end{defn}
Two examples of this coproduct, taken from QED, are:
\begin{align*}
\Delta(
\parbox{25pt}{\begin{fmfgraph*}(25,11)
      \fmfleft{l}
      \fmfright{r}
      \fmf{plain}{l,v1,v2,v3,v4,r}
      \fmf{photon,left,tension=0}{v1,v4}
      \fmf{photon,left,tension=0}{v2,v3}
\end{fmfgraph*}}
) &= 
\parbox{25pt}{\begin{fmfgraph*}(25,11)
      \fmfleft{l}
      \fmfright{r}
      \fmf{plain}{l,v1,v2,v3,v4,r}
      \fmf{photon,left,tension=0}{v1,v4}
      \fmf{photon,left,tension=0}{v2,v3}
\end{fmfgraph*}}
 \otimes 1 + 1 \otimes 
\parbox{25pt}{\begin{fmfgraph*}(25,11)
      \fmfleft{l}
      \fmfright{r}
      \fmf{plain}{l,v1,v2,v3,v4,r}
      \fmf{photon,left,tension=0}{v1,v4}
      \fmf{photon,left,tension=0}{v2,v3}
\end{fmfgraph*}}
+
\parbox{25pt}{
\begin{fmfgraph*}(25,11)
      \fmfleft{l}
      \fmfright{r}
      \fmf{plain}{l,v2,v3,r}
      \fmf{photon,left,tension=0}{v2,v3}
\end{fmfgraph*}}
\otimes
\parbox{25pt}{
\begin{fmfgraph*}(25,11)
      \fmfleft{l}
      \fmfright{r}
      \fmf{plain}{l,v1,v4,r}
      \fmf{photon,left,tension=0}{v1,v4}
\end{fmfgraph*}}
+
\parbox{25pt}{
\begin{fmfgraph*}(25,11)
      \fmfleft{l}
      \fmfright{r}
      \fmf{plain}{l,v2,v3,r}
      \fmf{photon,left,tension=0}{v2,v3}
\end{fmfgraph*}}
{ }_\bullet
\otimes
\parbox{25pt}{
\begin{fmfgraph*}(25,11)
      \fmfleft{l}
      \fmfright{r}
      \fmf{plain}{l,v1,v3,v4,r}
      \fmfdot{v3}
      \fmf{photon,left,tension=0}{v1,v4}
\end{fmfgraph*}}~,
\\[3mm]
\Delta(
  \parbox{25pt}{
    \begin{fmfgraph*}(25,25)
      \fmfleft{l}
      \fmfright{r}
      \fmf{phantom}{l,v1,v2,r}
      \fmf{photon}{l,v1}
      \fmf{photon}{v2,r}
      \fmf{phantom,left,tension=0,tag=1}{v1,v2}
      \fmf{phantom,right,tension=0,tag=2}{v1,v2}
      \fmffreeze
      \fmfi{plain}{subpath (0,.8) of vpath1(__v1,__v2)}
      \fmfi{plain}{subpath (0,.8) of vpath2(__v1,__v2)}
      \fmfi{plain}{subpath (0.8,1.2) of vpath1(__v1,__v2)}
      \fmfi{plain}{subpath (0.8,1.2) of vpath2(__v1,__v2)}
      \fmfi{photon}{point .8 of vpath1(__v1,__v2) .. point .8 of vpath2(__v1,__v2)}
      \fmfi{photon}{point 1.2 of vpath1(__v1,__v2) .. point 1.2 of vpath2(__v1,__v2)}
      \fmfi{plain}{subpath (1.2,2) of vpath1(__v1,__v2)}
      \fmfi{plain}{subpath (1.2,2) of vpath2(__v1,__v2)}
    \end{fmfgraph*}}
) &= 
 \parbox{25pt}{
    \begin{fmfgraph*}(25,25)
      \fmfleft{l}
      \fmfright{r}
      \fmf{phantom}{l,v1,v2,r}
      \fmf{photon}{l,v1}
      \fmf{photon}{v2,r}
      \fmf{phantom,left,tension=0,tag=1}{v1,v2}
      \fmf{phantom,right,tension=0,tag=2}{v1,v2}
      \fmffreeze
      \fmfi{plain}{subpath (0,.8) of vpath1(__v1,__v2)}
      \fmfi{plain}{subpath (0,.8) of vpath2(__v1,__v2)}
      \fmfi{plain}{subpath (0.8,1.2) of vpath1(__v1,__v2)}
      \fmfi{plain}{subpath (0.8,1.2) of vpath2(__v1,__v2)}
      \fmfi{photon}{point .8 of vpath1(__v1,__v2) .. point .8 of vpath2(__v1,__v2)}
      \fmfi{photon}{point 1.2 of vpath1(__v1,__v2) .. point 1.2 of vpath2(__v1,__v2)}
      \fmfi{plain}{subpath (1.2,2) of vpath1(__v1,__v2)}
      \fmfi{plain}{subpath (1.2,2) of vpath2(__v1,__v2)}
    \end{fmfgraph*}}
 \otimes 1 + 1 \otimes 
\parbox{25pt}{
    \begin{fmfgraph*}(25,25)
      \fmfleft{l}
      \fmfright{r}
      \fmf{phantom}{l,v1,v2,r}
      \fmf{photon}{l,v1}
      \fmf{photon}{v2,r}
      \fmf{phantom,left,tension=0,tag=1}{v1,v2}
      \fmf{phantom,right,tension=0,tag=2}{v1,v2}
      \fmffreeze
      \fmfi{plain}{subpath (0,.8) of vpath1(__v1,__v2)}
      \fmfi{plain}{subpath (0,.8) of vpath2(__v1,__v2)}
      \fmfi{plain}{subpath (0.8,1.2) of vpath1(__v1,__v2)}
      \fmfi{plain}{subpath (0.8,1.2) of vpath2(__v1,__v2)}
      \fmfi{photon}{point .8 of vpath1(__v1,__v2) .. point .8 of vpath2(__v1,__v2)}
      \fmfi{photon}{point 1.2 of vpath1(__v1,__v2) .. point 1.2 of vpath2(__v1,__v2)}
      \fmfi{plain}{subpath (1.2,2) of vpath1(__v1,__v2)}
      \fmfi{plain}{subpath (1.2,2) of vpath2(__v1,__v2)}
    \end{fmfgraph*}}
+2~ \parbox{25pt}{
  \begin{fmfgraph*}(25,25)
      \fmfleft{l}
      \fmfright{r1,r2}
      \fmf{photon}{l,v}
      \fmf{plain}{v,v1,v3,r1}
      \fmf{plain}{v,v2,v4,r2}
      \fmffreeze
      \fmf{photon}{v1,v2}      
      \fmf{photon}{v3,v4}
  \end{fmfgraph*}}
\otimes 
\parbox{25pt}{
\begin{fmfgraph*}(25,25)
  \fmfleft{l}
  \fmfright{r}
  \fmf{phantom}{l,v1,v2,r}
  \fmf{photon}{l,v1}
  \fmf{photon}{v2,r}
  \fmf{plain,left,tension=0}{v1,v2}
  \fmf{plain,left,tension=0}{v2,v1}
\end{fmfgraph*}}
+2~ \parbox{25pt}{
  \begin{fmfgraph*}(25,25)
      \fmfleft{l}
      \fmfright{r1,r2}
      \fmf{photon}{l,v}
      \fmf{plain}{v,v1,r1}
      \fmf{plain}{v,v2,r2}
      \fmffreeze
      \fmf{photon}{v1,v2}
  \end{fmfgraph*}}
\otimes 
\parbox{25pt}{
    \begin{fmfgraph*}(25,25)
      \fmfleft{l}
      \fmfright{r}
      \fmf{phantom}{l,v1,v2,r}
      \fmf{photon}{l,v1}
      \fmf{photon}{v2,r}
      \fmf{phantom,left,tension=0,tag=1}{v1,v2}
      \fmf{phantom,right,tension=0,tag=2}{v1,v2}
      \fmffreeze
      \fmfi{plain}{subpath (0,1) of vpath1(__v1,__v2)}
      \fmfi{plain}{subpath (0,1) of vpath2(__v1,__v2)}
      \fmfi{photon}{point 1 of vpath1(__v1,__v2) .. point 1 of vpath2(__v1,__v2)}
      \fmfi{plain}{subpath (1,2) of vpath1(__v1,__v2)}
      \fmfi{plain}{subpath (1,2) of vpath2(__v1,__v2)}
    \end{fmfgraph*}
}+  \parbox{25pt}{
  \begin{fmfgraph*}(25,25)
      \fmfleft{l}
      \fmfright{r1,r2}
      \fmf{photon}{l,v}
      \fmf{plain}{v,v1,r1}
      \fmf{plain}{v,v2,r2}
      \fmffreeze
      \fmf{photon}{v1,v2}
  \end{fmfgraph*}}
 \parbox{25pt}{
  \begin{fmfgraph*}(25,25)
      \fmfleft{l}
      \fmfright{r1,r2}
      \fmf{photon}{l,v}
      \fmf{plain}{v,v1,r1}
      \fmf{plain}{v,v2,r2}
      \fmffreeze
      \fmf{photon}{v1,v2}
  \end{fmfgraph*}}
\otimes  
\parbox{25pt}{
\begin{fmfgraph*}(25,25)
  \fmfleft{l}
  \fmfright{r}
  \fmf{phantom}{l,v1,v2,r}
  \fmf{photon}{l,v1}
  \fmf{photon}{v2,r}
  \fmf{plain,left,tension=0}{v1,v2}
  \fmf{plain,left,tension=0}{v2,v1}
\end{fmfgraph*}}~.
\end{align*}

\bigskip

The above Hopf algebra is an example of a connected graded Hopf algebra: it is graded by the {\it loop number $L(\Gamma)$} of a graph $\Gamma$. Indeed, one checks that the coproduct (and obviously also the product) satisfy the grading by loop number and $H^0$ consists of complex multiples of the empty graph, which is the unit in $H$, so that $H^0=\C 1$. We denote by $q_l$ the projection in $H$ onto $H^l$. 

In addition, there is another grading on this Hopf algebra. It is given by the number of vertices and already appeared in \cite{CK99}. However, since we consider vertices and edges of different types (wiggly, dotted, straight, {\it et cetera}), we extend to a multigrading as follows. As in \cite{Sui07}, we denote by $m_{\Gamma, r}$ the number of vertices/internal edges of type $r$ appearing in $\Gamma$, for $r \in R$. Moreover, let $n_{\gamma,r}$ be the number of connected components of $\gamma$ with residue $r$. For each $v \in R_V$ we define a degree $d_v$ by setting
$$
d_v(\Gamma) = m_{\Gamma,v} - n_{\Gamma,v}.
$$
The multidegree indexed by $R_V$ is compatible with the Hopf algebra structure as follows easily from the following relation:
\begin{align*}
m_{\Gamma/\gamma,v} = m_{\Gamma,v} - m_{\gamma,v} + n_{\gamma,v},
\end{align*}
and the fact that $m_{\Gamma\Gamma',v} = m_{\Gamma,v}+m_{\Gamma',v}$, and $n_{\Gamma\Gamma',v} = n_{\Gamma,v}+n_{\Gamma',v}$. This gives a decomposition 
$$
H= \bigoplus_{(n_1,\ldots,n_k) \in \Z^k} H^{n_1,\ldots,n_k}, 
$$
where $k=|R_V|$. We denote by $p_{n_1,\ldots,n_k}$ the projection onto $H^{n_1,\ldots,n_k}$. Note that also $H^{0,\cdots, 0}=\C 1$.

\begin{lma}
\label{lma:rel-degrees}
There is the following relation between the grading by loop number and the multigrading by number of vertices:
$$
\sum_{v \in R_V} (N(v)-2)d_v = 2 L
$$
where $N(v)$ is the valence of the vertex $v$.
\end{lma}

\begin{proof}
This can be easily proved by induction on the number of internal edges using invariance of the quantity $\sum_v (N(v)-2)d_v - 2 L$ under the adjoint of an edge.
\end{proof}

The group $\Hom_\C(H,\C)$ dual to $H$ is called the {\it group of diffeographism}. This name was coined in \cite{CK00} motivated by its relation with the group of (formal) diffeomorphisms of $\C$, whose definition we recall in the next section. Stated more precisely, they constructed a map from the group of diffeographism to the group of formal diffeomorphisms. We will establish this result in general ({i.e.} for any quantum field theory) in Section \ref{sect:setup} below.

\subsection{Formal diffeomorphisms}
Another Hopf algebra that will be of interest is that dual to the group $\Diff(\C,0)$ of formal diffeomorphisms of $\C$ tangent to the identity, it is known in the literature as the Fa\`a di Bruno Hopf algebra (see for instance the short review \cite{FGV05}).
The elements of this group are given by formal power series: 
\begin{equation}
\label{eq:formal-diffeo}
f(x) = x \sum_{n\geq 0} a_n(f) x^n ; \qquad a_0(f)=1
\end{equation}
with the composition law given by $(f \circ g)(x) = f(g(x))$. 
The coordinates $\{a_n\}$ generate a Hopf algebra with the coproduct, counit and antipode defined in terms of the pairing $\langle a_n, f\rangle := a_n(f)$ as
\begin{align}
\label{hopf-diff}
\langle \Delta( a_n) , f \otimes g \rangle = \langle a_n , g \circ f \rangle. \qquad
\epsilon(a_n) = \langle a_n ,1\rangle, \qquad
\langle S(a_n), f \rangle = \langle a_n, f^{-1} \rangle 
\end{align}

A convenient expression for the coproduct on $a_n$ can be given as follows \cite{BFK06}. Consider the generating series
$$ 
A(x)= x \sum_{n\geq 0} a_n x^n; \qquad a_0 = 1
$$
where $x$ is considered as a formal parameter. Then the coproduct can be written as
\begin{align}
\label{eq:cop-res}
\Delta A(x) = \sum_{n \geq 0} A(x)^{n+1} \otimes a_n
\end{align}
One readily checks that indeed $\langle \Delta A(x), g \otimes f \rangle = f(g(x))$.

\begin{rem}
Actually, this Hopf algebra is the dual of the opposite group of $\Diff(\C,0)$. Instead of acting on $\C$ as formal diffeomorphisms, the opposite group $\Diff(\C,0)^\op$ can be characterized by its action on the algebra $\C[[x]]$ of formal power series in $x$. On the generator $x$, the action of $\Diff(\C,0)^\op$ is defined by the same formula \eqref{eq:formal-diffeo} but it is extended to all of $\C[[x]]$ as an algebra map. We will denote in the following this group by $\Aut_1(\C[[x]]):=\Diff(\C,0)^\op$.
\end{rem}

Clearly, we have an analogous definition of formal diffeomorphisms of $\C^k$ tangent to the identity. The group $\Diff(\C^k,0)$ consists of elements:
$$
f(x) = \big( f_1(x ),\ldots, f_k(x) \big)
$$
where each $f_i$ is a formal power series of the following form 
$$f_i(x ) = x_i(\sum a^{(i)}_{n_1\cdots n_k}(f) x_1^{n_1} \cdots x_k^{n_k})
$$ 
with $a^{(i)}_{0,\ldots,0}=1$ and $x=(x_1, \cdots ,x_k)$.

Again, there is a dual Hopf algebra generated by the coordinates $a^{(i)}_{n_1\cdots n_k}$ with the coproduct, counit and antipode defined by the analogous formula to Eq. \eqref{hopf-diff}. 
\begin{lma}
\label{lma:cop-diff-k}
On the generating series $A_i(x) = x_i(\sum a^{(i)}_{n_1\cdots n_k}  x_1^{n_1} \cdots x_k^{n_k})$ the coproduct equals
$$
\Delta(A_i(x)) = \sum_{n_1, \ldots, n_k}  A_i(x) \left( A_1(x) \right)^{n_1} \cdots \left( A_k(x) \right)^{n_k} \otimes a^{(i)}_{n_1\cdots n_k}.
$$
\end{lma}

Closely related to these groups of formal diffeomorphisms, is the group of invertible power series in $k$ parameters, denoted $\C[[x_1,\ldots, x_k]]^\times$. As above, it consists of formal series $f$ with non-vanishing first coefficient $a_0(f)\neq 0$, but with product given by the algebra multiplication. The formula for the inverse is given by the Lagrange inversion formula for formal power series.

\subsection{Birkhoff decomposition}
\label{sect:birkhoff}
We now briefly recall how renormalization is an instance of a Birkhoff decomposition in the group of characters of $H$ as established in \cite{CK99}. Let us first recall the definition of a Birkhoff decomposition.

We let $\gamma: C \to G$ be a loop with values in an arbitrary complex Lie group $G$, defined on a smooth simple curve $C \subset \P_1(\C)$. Let $C_\pm$ be the two complements of $C$ in $\P_1(\C)$, with $\infty \in C_-$. A {\it Birkhoff decomposition} of $\gamma$  is a factorization of the form 
$$
\gamma(z) = \gamma_-(z)^{-1} \gamma_+(z); \qquad (z \in C),
$$
where $\gamma_\pm$ are (boundary values of) two holomorphic maps on $C_\pm$, respectively, with values in $G$. This decomposition gives a natural way to extract finite values from a divergent expression. Indeed, although $\gamma(z)$ might not holomorphically extend to $C_+$, $\gamma_+(z)$ is clearly finite as $z\to 0$.

Now consider a Feynman graph $\Gamma$ in the Hopf algebra $H$. Via the so-called Feynman rules -- which are dictated by the Lagrangian of the theory -- one associates to $\Gamma$ the Feynman amplitude $U(\Gamma)(z)$. It depends on some regularization parameter, which in the present case is a complex number $z$ (dimensional regularization). The famous divergences of quantum field theory are now `under control' and appear as poles in the Laurent series expansion of $U(\Gamma)(z)$. 

On a curve around $0 \in \P^1(\C)$ we can define a loop $\gamma$ by $\gamma(z)(\Gamma):=U(\Gamma)(z)$ which takes values in the group of diffeographisms $G=\Hom_\C(H,\C)$. Connes and Kreimer proved the following general result in \cite{CK99}.
\begin{thm}
Let $H$ be a graded connected commutative Hopf algebra with character group $G$. Then any loop $\gamma: C \to G$ admits a Birkhoff decomposition.
\end{thm}
In fact, an explicit decomposition can be given in terms of the group $G(K)= \Hom_\C(H,K)$ of $K$-valued characters of $H$, where $K$ is the field of convergent Laurent series in $z$.\footnote{In the language of algebraic geometry, there is an affine group scheme $G$ represented by $H$ in the category of commutative algebras. In other words, $G=\Hom_\C(H,~ . ~)$ and $G(K)$ are the $K$-points of the group scheme. } 
If one applies this to the above loop associated to the Feynman rules, the decomposition gives exactly renormalization of the Feynman amplitude $U(\Gamma)$: the map $\gamma_+$ gives the renormalized Feynman amplitude and the $\gamma_-$ provides the counterterm. 

\bigskip

Although the above construction gives a very nice geometrical description of the process of renormalization, it is a bit unphysical in that it relies on individual graphs that generate the Hopf algebra. Rather, as mentioned before, in physics the probability amplitudes are computed from the full expansion of Green's functions. Individual graphs do not correspond to physical processes and therefore a natural question to pose is how the Hopf algebra structure behaves at the level of the Green's functions. We will see in the next section that they generate Hopf subalgebras, {i.e.} the coproduct closes on Green's functions. Here the so-called Slavnov--Taylor identities for the couplings will play a prominent role.

\section{Feynman graphs and formal diffeomorphisms}
\label{sect:setup}
In this section, the group of formal diffeomorphisms of $\C$ will be shown to arise as a quotient of the group of diffeographisms. As before, it is very convenient to work in a dual manner with the relevant Hopf algebras. 

We define the {\it 1PI Green's functions} by
\begin{equation*}
G^e = 1 - \sum_{\res(\Gamma)=e} \frac{\Gamma}{\Sym(\Gamma)},\qquad G^{v} = 1 + \sum_{\res(\Gamma)=v} \frac{\Gamma}{\Sym(\Gamma)} 
\end{equation*}
with $e \in R_E, v \in R_V$. The restriction of the sum to graphs $\Gamma$ at loop order $L(\Gamma)=l$ is denoted by $G^r_l$.

The following prepares for renormalization in the BV-formalism, which differs slightly from the usual wave function and coupling constant renormalization (see for instance \cite[Section 6]{Ans94}). For each $\phi \in \Phi$ we assume that we are given elements $C^\phi \in H$ such that the following hold:
\begin{enumerate}
\item If $\phi$ only appears linearly in the Lagrangian then $C^\phi C^{\phi_{i_1}} \cdots C^{\phi_{i_1}}=1$ for $\iota(v) \propto \phi \phi_{i_1} \cdots \phi_{i_m}$.
\item If $\iota(e) \propto \phi \bar\phi$ then $C^\phi C^{\bar\phi} = G^e$.
\item For any field $\phi_i$ we have $C^{K_{\phi_i}} C^{\phi_i} =1$.
\end{enumerate}
Note that in general the $C^\phi$'s are not uniquely determined from these conditions. However, in theories of interest such as Yang--Mills gauge theories, they actually are as illustrated by the next example.
\begin{ex}
For pure Yang--Mills gauge theories (see for notation Example \ref{ex:gauge} below) we have 
\begin{gather*}
C^A = \sqrt{G^\glu}; \quad C^\omega = (G^\gho) \sqrt{ G^\glu}; \quad C^{\bar\omega} = (G^\glu)^{-\half}; \quad C^h = (G^\glu)^{-\half},
\end{gather*}
and $C^{K_\phi} = (C^\phi)^{-1}$ for $\phi=A,\omega,\bar\omega,h$. Note that $C^\omega C^{\bar\omega} = G^{\gho}$ which -- as we shall see in Section \ref{sect:ym} below -- will be the usual wave function renormalization for the ghost propagator. 
\end{ex}
Returning to the general setup, we assume that we have defined such elements $C^\phi$ for all $\phi \in \Phi$.

\begin{rem}
Let us pause to explain the meaning of the inverse of Green's functions in our Hopf algebra. Since any Green's function $G^r$ for $r \in R$ starts with the identity, we can surely write its inverse formally as a geometric series. Recall that the Hopf algebra is graded by loop number. Hence, the inverse of a Green's function at a fixed loop order is in fact well-defined; it is given by restricting the above formal series expansion to this loop order. More generally, we understand any real power of a Green's function in this manner.
\end{rem}

In earlier work \cite[Eq. (11)]{Sui07c}, we have shown that the coproduct on Green's functions takes the following form:
\begin{equation}
\label{cop-green}
\Delta(G^r) = G^r \otimes 1 + G^r \sum_{\res(\Gamma)=r}~ \prod_{v\in R_V, v\neq r} \left( \frac{G^v}{\prod_\phi \left(C^\phi\right)^{N_\phi(v)} } \right)^{m_{\Gamma,v}} \otimes \frac{\Gamma}{\Sym(\Gamma)}.
\end{equation}
Here $N_\phi(r)$ is the number of lines corresponding to the field $\phi \in \Phi$ attached to $r \in R$; clearly, the total number of lines attached to $r$ can be written as $N(r)=\sum_{\phi \in \Phi} N_\phi(r)$. 

\begin{rem}
In order to reduce the above formula to Eq. (11) in \cite{Sui07c} one observes that if $v$ does not involve a BRST-source term then
$$
\frac{G^v}{\prod_\phi \left(C^\phi\right)^{N_\phi(v)} }=\frac{G^v}{\prod_{e \in R_E} \left(G^e\right)^{N_e(v)/2} }
$$
since a fermionic field $\phi$ will always be accompanied by the field $\bar\phi$ on a vertex that does not involve a BRST-source (cf. Remark \ref{rem:psipsibar}), thus reducing the above formula to Eq. (11) in {\it loc. cit.}. It is sufficient to consider only the case of no BRST-sources since in either case (for $r$ with or without BRST-source) the $v$'s appearing in the above formula will never involve a BRST-source. 
\end{rem}

\begin{prop}
\label{prop:cop-Yv}
Define elements $Y_v \in H$ for $v \in R_V$ as formal expansions:
$$
Y_v := \frac{G^v}{\prod_\phi \left(C^\phi\right)^{N_\phi(v)} }.
$$
The coproduct on $(Y_v)^\alpha$ with $\alpha \in \R$ is given by
$$
\Delta(Y_v^\alpha) = \sum_{n_1 \cdots n_k} Y_v^\alpha Y_{v_1}^{n_1} \cdots Y_{v_k}^{n_k}  \otimes p_{n_1 \cdots n_k} (Y_v^\alpha),
$$
where $p_{n_1 \cdots n_k}$ is the projection onto graphs containing $n_i$ vertices of the type $v_i$ ($i=1,\ldots,k=|R_V|$). 
\end{prop}
\begin{proof}
First, one can obtain from Eq. \eqref{cop-green} the coproduct on $G^r$ as
\begin{align}
\Delta(G^r) &=\sum_{n_1, \ldots, n_k } G^r Y_{v_1}^{n_1} \cdots Y_{v_k}^{n_k} \otimes p_{n_1,\ldots,n_k} (G^r)\nonumber
\intertext{
which holds for any $r \in R$. A long but straightforward computation involving formal power series expansions yields the following expression for real powers (in the above sense) of the Green's functions:}
\label{cop-Ge}
\Delta ( (G^r )^\alpha ) &= \sum_{n_1, \ldots, n_k } (G^r)^\alpha  Y_{v_1}^{n_1} \cdots Y_{v_k}^{n_k}
 \otimes p_{n_1,\ldots,n_k} ((G^r)^\alpha),
\intertext{for $r \in R$ and $\alpha \in \R$. Thus, also}
\Delta ( (C^\phi )^\alpha ) &= \sum_{n_1, \ldots, n_k } (C^\phi)^\alpha  Y_{v_1}^{n_1} \cdots Y_{v_k}^{n_k}
 \otimes p_{n_1,\ldots,n_k} ((C^\phi)^\alpha),
\label{cop-Cphi}
\end{align}
and a combination of these formulas together with the fact that $\Delta$ is an algebra map yields the desired cancellations so as to obtain the stated formula.
\end{proof}

\begin{rem}
\label{rem:XY}
In \cite{Sui07b,Sui07c} we considered the elements $X_v := (Y_v)^{1/(N(v)-2)}$ for vertices $v$ of valence greater than 2. Currently, we are including vertices of valence 2 to incorporate mass terms, which  motivates the definition of $Y_v$ instead. 
\end{rem}

There is a striking similarity between the above formula for $\Delta(Y_v)$ and the coproduct in the Hopf algebra dual to $\Diff(\C^k,0)$, as in Lemma \ref{lma:cop-diff-k}. In fact, we have the following
\begin{corl}
\label{corl:graph-diffeo}
There is a surjective map from the Hopf algebra dual to the group $\Diff(\C^k,0)^\op$ to the Hopf subalgebra in $H$ generated by $p_{n_1\cdots n_k} (Y_v)$.
\end{corl}
\begin{proof}
Whenever $(n_1, \ldots, n_k) \neq (0,\ldots,0)$, we map the coordinates $a_{n_1 \ldots , n_k}^{(i)}$ of $\Diff(\C^k,0)$ to the elements $p_{n_1, \ldots , n_i -1, \ldots, n_k}(Y_{v_i}) \in H$, with $k=|R_V|$. Indeed, $p_{n_1 \cdots n_k}(Y_{v_i})$ vanishes for all $n_j < 0$ ($j\neq i$) and $n_i < -1$, explaining the shift in the $i$-th index. Moreover, both $a^{(i)}_{0,\ldots, 0}$ and $p_{0, \ldots ,0}(Y_{v_i})$ are equal to the identity.
\end{proof}

Actually, with Equation \eqref{cop-Ge} above it is easy to see that the algebra generated by $p_{n_1\cdots n_k} (Y_v)$ and $p_{n_1\cdots n_k} (G^e)$ for $v \in R_V$ and $e\in R_E$ is a Hopf subalgebra, which we denote by $H_R$. Equivalently, we can take as generators for $H_R$ the elements $p_{n_1\cdots n_k} (Y_v)$ and $p_{n_1\cdots n_k} (C^\phi)$.
In Proposition \ref{prop:action} below we will show that the corresponding dual group is in fact a subgroup of the semi-direct product $(\C[[x_1, \ldots, x_k]]^\times)^{|R_E|} \rtimes \Diff(\C^k,0)$.

\bigskip

We will next establish that a quotient of the Hopf algebra generated by $p_{n_1, \ldots, n_k}(Y_v)$ by a certain Hopf ideal is isomorphic to the Hopf algebra dual to (a subgroup of) $\Aut_1(\C[[x]]) \equiv \Diff(\C,0)^\op$. The latter is indeed a subgroup of $\Diff(\C^k,0)^\op$ under the diagonal embedding.  

\begin{thm}{\cite{Sui07c}}
\label{thm:hopfideal}
The ideal $J'$ in $H_R$ generated by $q_l\left(Y_{v'}^{N(v)-2} - Y_{v}^{N(v')-2}  \right)$ for $v',v \in R_V$ of valence greater than 2 ($l \geq 0$), and $Y_v$ for all $v$ of valence 2 is a Hopf ideal, {i.e.}
$$
\Delta(J') \subset J' \otimes H_R + H_R \otimes J'. 
$$
\end{thm}
\begin{proof}
First of all, with Proposition \ref{prop:cop-Yv}, the coproduct on $Y_v$ for $\val(v)=2$ is readily found to be an element in $J' \otimes H_R + H_R \otimes J'$.
With Proposition \ref{prop:cop-Yv}, we can write the coproduct on the other generators of $J'$ as
\begin{multline*}
\Delta \left( Y^{N(v)-2}_{v'} - Y^{N(v')-2}_{v}\right) = 
 \sum_{n}  Y^{N(v')-2}_{v}  Y_{v_1}^{n_1 } \cdots Y_{v_k}^{n_k }  \otimes p_{n} \big( Y^{N(v)-2}_{v'} - Y^{N(v')-2}_{v}\big)\\
+\sum_{n} \bigg[Y^{N(v)-2}_{v'} - Y^{N(v')-2}_{v} \bigg]  Y_{v_1}^{n_1 } \cdots Y_{v_k}^{n_k }  \otimes p_{n} \left( Y^{N(v)-2}_{v'} \right)
\end{multline*}
with $n$ the multi-index $(n_1,\ldots,n_k)$.
The second term is clearly an element in $J' \otimes H_R$. For the first term, note that each $n_i$'th power of $Y_{v_i}$ can be written as
$$
Y_{v_i}^{n_i} = Y_{v_i}^{n_i  \frac{N(v)-2}{N(v)-2}} = Y_{v}^{n_i  \frac{N(v_i)-2}{N(v)-2}} + J'.
$$
Hence, the first term becomes {\it modulo} $J' \otimes H_R$
$$
\sum_{n_1 \cdots n_k}  \left( Y^{1/N(v)-2}_{v} \right)^{n_1 (N(v_1)-2) + \cdots + n_k (N(v_k)-2) } 
\otimes p_{n_1 \cdots n_k} \bigg( Y^{N(v)-2}_{v'} - Y^{N(v')-2}_{v}\bigg).
$$
Appealing to Lemma \ref{lma:rel-degrees} now allows us to write this in terms of the loop number $l$ to finally obtain for the first term 
$$
\sum_{l=0}^\infty Y_{v}^{\frac{2l+1}{N(v)-2)}} \otimes q_l \bigg(  Y^{N(v)-2}_{v'} - Y^{N(v')-2}_{v}\bigg).
$$
which is indeed an element in $H_R \otimes J'$. 
\end{proof}
As a consequence, the quotient Hopf algebra $\tilde H_R = H_R/J'$ is well-defined. In $\tilde H_R$ the relations  $Y_v^{N(v')-2} = Y_{v'}^{N(v)-2}$ are satisfied, or, in terms of the $X_v$ of Remark \ref{rem:XY} they are simply $X_{v'}=X_v$. In physics these identities are called {\it Slavnov--Taylor identities} for the couplings; we will see later how they appear naturally from the relations between coupling constants. Moreover, the fact that we put $Y_v = 0$ for vertices of valence 2 means that we consider a massless theory. In $\tilde H_R$ we can drop the subscript $v$ and use the notation $X :=  Y_v^{1/N(v)-2} \equiv X_v$ independent of $v \in R_V$ as long as $\val(v) >2$.

\begin{thm}
The coproduct in $\tilde H_R$ takes the following form on the element $X$: 
$$
\Delta(X) = \sum_{l=0}^\infty \left( X \right)^{2l+1} \otimes q_l(X). 
$$
where $q_l$ is the projection in $\tilde H_R$ onto graphs of loop number $l$. 
\end{thm}
\begin{proof}
This follows directly by substituting $X$ for $X_v$ in the expression for $\Delta(X_v)$ in Proposition \ref{prop:cop-Yv} and using the relation from Lemma \ref{lma:rel-degrees} between the number of vertices and the loop number. 
\end{proof}
Thus, the Hopf algebra $\tilde H_R$ contains a Hopf subalgebra that is generated by $q_l(X)$ and a comparison with Eq. \eqref{eq:cop-res} yields -- after identifying $q_l(X)$ with $a_{2l}$ -- the following result.
\begin{thm}
The graded Hopf subalgebra in $\tilde H_R$ generated by $q_l\left(X\right)$ for $l=0,1,\ldots$ is isomorphic to the Hopf algebra of the group of odd formal diffeomorphisms of $\C$ tangent to the identity. 
In other words, there is a homomorphism from the group of diffeographisms to $\Diff(\C,0)^\op\equiv \Aut_1(\C[[x]])$.
\end{thm}
This generalizes the result of \cite{CK00} where such a map was constructed explicitly in the case of (massless) $\phi^3$-theory; for other theories a map has been constructed by Cartier and Krajewski. In the next section, we will explore its relation with the group of formal diffeomorphisms acting on the space of coupling constants. 

\section{Coaction on the fields and coupling constants}
\label{sect:coaction}
In this section, we will establish a connection between the Hopf algebra of Feynman graphs defined above and the fields, coupling constants and masses that characterize the field theory. This allows us to derive the Hopf ideals encountered in the previous section from the so-called master equation satisfied by the Lagrangian. 
Let us start by a careful setup of the algebra of local functions and functionals in the fields that constitute the field theory. Readers already familiar with this might want to skip to Section \ref{subsect:comodule} where the connection is established between the BV-algebra of fields and the renormalization Hopf algebras.

\subsection{Fields and BRST-sources}
Although we have already introduced the set of fields $\Phi$ above, we have not said precisely what we mean with a field. Let us do so in a bit more generality than needed. A {\it field} $\phi$ is a section of a vector bundle $E \to M$ on the background manifold $M$. If the rank of the vector bundle $E$ is $r$, the field is said to have $r$ components, in which case we can write locally $\phi = \phi^a e_a$ in terms of a basis $e_a$ of $E$. 
\begin{ex}
If $E= M \times \C$, then a section $\phi$ is a complex scalar field $\phi: M \to \C$; it has one component. 
\end{ex}
 
\begin{ex}
\label{ex:gauge}
Gauge fields are sections $A$ of $E = \Lambda^1 \otimes (P \times_G \g)$ with $P$ a $G$-principal bundle and $\g = \textup{Lie}(G)$. In the case that $P$ is trivial, this becomes a $\g$-valued one-form on $M$, {i.e.} $A$ is a section of $\Lambda^1(\g)$. In this case, the rank of the vector bundle is $\dim(M)\cdot \rank(\g)$ which leads to the familiar decomposition 
$$
A =  A^a_\mu dx^\mu T^a,
$$ 
with $\{ T^1, \ldots, T^{\rank(\g)} \}$ a basis for $\g$ and summation is understood.
\end{ex}

If we consider a set $\Phi$ consisting of $2N$ fields, we have specified $2N$ (graded) vector bundles each of which has a corresponding field as its section. As said, we will assume that the fields come in pairs of a field $\phi_i$ and an BRST-source $K_{\phi_i}$ ($i=1,\ldots, N$) and we write $E_i$ and $E_i^\vee$ for the corresponding vector bundles which are of equal rank. In fact, $E_i^\vee$ is the dual vector bundle of $E_i$, although shifted in degree as we make more precise now. The fields $\phi_i$ are understood to have a so-called {\it ghost degree} $\gh(\phi_i) \in \Z$ which is then extended to the BRST-sources by
$$
\gh(K_{\phi_i}) := -\gh(\phi_i) - 1.
$$
In the physics literature, this is usually called the (total) {\it ghost number}. Summarizing, the elements of $\Phi$ constitute a section of the total vector bundle $E_\tot$:
$$
(\phi_1,K_{\phi_1},\ldots , \phi_N,K_{\phi_N}) : M \to E_\tot= \bigoplus_{i=1}^N E_i \oplus  E_i^\vee,
$$
The grading on the fields turn $E_\tot$ into a graded vector bundle. 

\begin{ex}
In Section \ref{sect:ym} below, we will focus on pure Yang-Mills gauge theories. In that case, there is the gauge field $A$ as in Example \ref{ex:gauge} which (in the trivial bundle case) is a section of $\Lambda^1 \otimes M \times \g$, {i.e.} an element of $\Omega^1(\g)$. The so-called ghost fields $\omega$ and $\bar\omega$ are assigned to each generator of $\g$, in components $\omega= \omega^a T^a$ and $\bar\omega=\bar\omega^a T^a$. Their ghost degrees are defined to be $1$ and $-1$, respectively, so that $\omega$ is a section in $\Omega^0(\g[-1])$ and $\bar\omega$ in $\Omega^0(\g[1])$. Also, there is the so-called auxiliary -- or Nakanishi--Lantrup -- field $h = h^a T^a$, which is a section in $\Omega^0(\g)$ and of degree $0$. 

Corresponding to these fields, there are the BRST-sources $K_A$, $K_\omega$, $K_{\bar\omega}$ and $K_h$ which are of respective ghost degree $-1$, $-2$, $0$ and $-1$. Thus, the field content of pure Yang-Mills gauge theories can be summarized by the following sections
\begin{align*}
(A,\omega, \bar\omega ,h)\in &~\Omega^1(\g) \oplus   \Omega^0( \g[-1]) \oplus \Omega^0 (\g[1]) \oplus \Omega^0(\g),\\
(K_A, K_\omega,K_{\bar\omega}, K_h) \in &~ \X( \g[1]) \oplus \Omega^0(\g) \oplus \Omega^0(\g[2])  \oplus \Omega^0(\g[1]),
\end{align*}
where $\X(\g)$ denotes $\g$-valued vector fields. Taken all together, they form a section of the total bundle. 
\end{ex}

\subsection{Jet bundles in Lagrangian field theory}
Let us now `prolong' this total bundle $E_\tot$ and construct the jet bundle $J^\infty (E_\tot)$. First, we generalize a little and briefly recall the theory of jet bundles. We refer to \cite{Sau89} for more details.

Let $\pi: E \to M$ be a vector bundle on an $m$-dimensional manifold $M$ and suppose $u \in \Gamma(M,E)$ is a smooth section. For each $x \in M$ consider a neighborhood $\U$ and a local trivialization $\pi^{-1} (\U)  \simeq \U \times \R^k$ with coordinates $x^\mu, u^a(x)$ with $\mu=1, \ldots, m$ and $a = 1,\ldots,k$. 
\begin{defn}
The {\rm first-order jet $j^1_x(\sigma)$ of a section $\sigma$ of $E$ at $x$} is the equivalence class of sections for the relation  
$$
\sigma \sim \sigma' \iff \sigma(x) = \sigma'(x), \quad \partial_i \sigma(x) =\partial_i \sigma'(x) \qquad (i=1,\ldots,m).
$$
for $\sigma' \in \Gamma(M,E)$.
\end{defn}
The set $J^1(E)$ of all such equivalence classes,
$$
J^1(E) = \bigcup_{\begin{smallmatrix} x \in M \\ \sigma \in \Gamma(M,E) \end{smallmatrix}} j^1_x(\sigma),
$$
carries the structure of a vector bundle over $M$ -- with projection map $\pi_1 : j^1_x(\sigma) \mapsto x$ -- and is called the {\it first-order jet bundle of $E$}. A local trivialization $\pi_1^{-1}(\U) \simeq \U \times \R^{k+ k m}$ is given in terms of the local coordinates $\{ u^{a_1} ,\partial_\mu u^{a_2} \}$. Besides the structure of a vector bundle over $M$, $J^1(E)$ is also a vector bundle over $E$, with projection map defined by $\pi_{r,0} :j_x^1(\sigma)  \mapsto \sigma(x)$.

If we apply this construction repeatedly to the jet bundle itself, we obtain the {\it n'th-order jet bundle} of $E$ as
$$
J^n (E) := \underbrace{J^1 ( \cdots (J^1(}_{n \text{ times }}   E)) \cdots ).
$$
In other words, $J^n(E)$ consists of equivalence classes of sections, which are identified when their values and the values of their partial derivatives up to order $n$ are equal. As a consequence, local coordinates on $J^n(E)$ are given by $\{ u^{a_1}, \partial_\mu u^{a_2} , \ldots, \partial_{\mu_1} \cdots \partial_{\mu_n} u^{a_n}  \}$. 

From the above construction, it is clear that we can define maps $\pi_{n,n'} : J^n(E) \to J^{n'}(E)$ for $1 \leq n' < n$, which can be extended to $n'=0$ and $n'=n$ if we identify $J^0(E)$ with $E$ and $\pi_{n,n}$ with the identity map on $J^n(E)$. The inverse limit of the resulting inverse system $(J^n(E), \pi_{n,n'})$ is called the infinite jet bundle and is denoted by $J^\infty(E)$. As an infinite-dimensional vector bundle it has coordinates $\{ u^{a_1} , \partial_{\mu} u^{a_2}, \partial_{\mu_1}\partial_{\mu_2} u^{a_3}, \ldots  \}$.

\bigskip

The jet bundle formalism is very convenient for specifying a field theory in terms of a Lagrangian. Indeed, such a function does not only depend on the fields, but also on their partial derivatives. Nevertheless, the condition of locality imposes an upper bound on the order of these partial derivatives, which motivates the following definition. 
\begin{defn}
\label{defn:localform}
A {\rm local form} $L(x,u^{(n)})$ is a pullback of a horizontal differential form on some finite jet bundle $J^{n}(E)$ to $J^\infty(E)$, {i.e.} an element in $\Omega^{\bullet,0}(J^\infty E)$. 
The {\rm algebra of local forms} is denoted by $\Loc(E)$.  
\end{defn}
Since $\Omega^{\bullet,0}(J^\infty E) \simeq \Omega^\bullet(M) \otimes C^\infty(J^\infty E) $, a local form is the tensor product of a differential form on $M$ with a smooth function in the coordinates $x^\mu$ and $\partial_{\mu_1} \cdots \partial_{\mu_{n'}} u^{a}$ ($0 \leq n' \leq n$) for some finite positive integer $n$. If the vector bundle $E$ carries a grading, $E = \oplus_q E^{(q)}$ the algebra $\Loc(E)$ becomes bigraded, $L \in \Loc(E)$ of bidegree $(p,q)$ if $L$ has degree $p$ as a differential form and ghost degree $q$. In this case, we write
$$
\Loc(E) = \bigoplus_{p \geq 0, q\in \Z} \Loc^{(p,q)} (E),
$$
and we have $\Loc^{(p,q)}(E) \simeq \Omega^p(M) \otimes_{C^\infty(M)} \Loc^{(0,q)}(E)$. 

In the case that $E=E_\tot$ so that the sections of $E$ constitute a set of fields $\Phi$ as above, we also write $\Loc(\Phi)$ instead of $\Loc(E)$ (and similarly $\Loc^{(p,q)}(\Phi)$) and with a slight abuse of notation $u^a \equiv \phi^a$ for $\phi \in \Phi$. We distinguish between sections and coordinates by writing explicitly the dependence of the second on the position on $M$. Thus, the components of a section $\sigma \in \Gamma(M,E_\tot)$ are given by
$$
\phi^a(x) = (\phi^a \circ \sigma)(x) \in \R
$$
with $\phi^a$ on the right-hand-side the fiber coordinates $u^a$ of the direct summand of $E_\tot$ corresponding to $\phi$. In a similar manner, we write for the coordinates of the higher order jet bundles $\partial_{\vec{\mu}} u^a \equiv \partial_{\vec\mu} \phi^a $ for a multi-index $\vec\mu = (\mu_1, \ldots, \mu_k)$. The previous correspondence between coordinates and sections generalizes to
$$
\partial_{\vec{\mu}} \phi^a  (x) = (\partial_{\vec{\mu}}\phi^a \circ j^\infty\sigma) (x).
$$
in terms of the infinite jet $j^\infty \sigma \in \Gamma(M,J^\infty (E_\tot))$ defined by the smooth section $\sigma \in \Gamma(M,E_\tot)$.

\begin{ex}
A scalar field theory is defined by the following Lagrangian $L \in \Loc^{(m,0)}(M \times \R)$:
$$
L(x,\phi, \partial_i \phi)=\half d \phi * d \phi - V(\phi) (*1)
$$
with $V(\phi)$ a polynomial in the field $\phi \in \Gamma(M \times \R)$. 
\end{ex}

To any Lagrangian $L$, defined in general as a local $m$-form of the fields ($m=\dim M$), one can associate the Lagrangian density $L(x, \phi(x)) := (j^\infty \sigma)^* L(x,\phi)$, evaluated at a section $\sigma$ of $E$. 
This density can be integrated to give the so-called {\it action}
$$
S[\phi] := \int_M L(x, \phi(x)).
$$
In general, we make the following definition. 
\begin{defn}
A {\rm local functional} $F[\phi]$ is the integral of the pullback of a local $m$-form, {i.e.} $F[\phi] =  \int_M L(x, \phi(x))$ for $L(x,\phi) \in \Loc^{(m,0)}(E)$. The free commutative algebra generated (over $\C$) by local functionals is denoted by $\F([E])$.
\end{defn}
Again, in the case that $E=E_\tot$ is associated to a set of fields $\Phi$ as above, we write $\F([\Phi])$ instead of $\F([E])$. The grading by ghost degree on local $m$-forms carries over to a grading on local functionals, which we also denote by $\gh(F)$ for $F \in \F([E])$.

\subsection{The anti-bracket}
\label{sect:BV}
We will now try to elucidate the above `doubling' of the fields (adding a BRST-source for every field) in terms of the structure of a {\it Gerstenhaber algebra} on the algebra of local functionals $\F([\Phi])$. 
Recall that a Gerstenhaber algebra \cite{Ger63} is a graded commutative algebra with a Lie bracket of degree 1 satisfying the graded Leibniz property:
$$
(x,yz) = (x,y)z + (-1)^{(|x|+1)|y|}y(x,z).
$$
Batalin and Vilkovisky encountered this structure in their study of quantum gauge theories \cite{BV81,BV83,BV84}. In fact, they invented what is now called a {\it BV-algebra} (see for instance \cite{Sta97}): a Gerstenhaber algebra with an additional operator $\tilde\Delta$ that satisfies:
$$
(x,y) = \tilde\Delta(xy) - \tilde\Delta(x) y + (-1)^{|x|} x \tilde\Delta(y). 
$$
We will define such an {\it anti-bracket} on the algebra of local functionals using the functional derivative. 
\begin{defn}
The left and right functional derivatives are the distributions defined by
$$
\frac{d}{dt} F[\phi + t \psi_\phi] = \int_M  \frac{\delta_L F} {\delta \phi^a(x)} \psi_\phi^a(x) d\mu(x)= \int_M \psi_\phi^a(x) \frac{\delta_R F} {\delta \phi^a(x) } d\mu(x),
$$
for test functions $\psi_\phi$ of the same ghost degree as $\phi \in \Phi$. 
\end{defn}
There is the following relation between the two functional derivatives:
$$
\frac{\delta_R F} {\delta \phi^a(x)}  = (-1)^{\gh(\phi) (\gh(F) - \gh(\phi))} \frac{\delta_L F} {\delta \phi^a(x)}.
$$
with $\gh$ the ghost degree.
\begin{prop}
The bracket $(\cdot,\cdot)$ defined by
\begin{equation*}
( F_1, F_2 ) = \sum_{i=1}^N \sum_{a=1}^{\rk E_i} \int_M \left[\frac{ \delta_R F_1}{\delta \phi^a_i(x)} \frac{ \delta_L F_2}{\delta K_{\phi_i}^a(x)} - \frac{ \delta_R F_1}{\delta K_{\phi_i}^a(x)} \frac{ \delta_L F_2}{\delta \phi^a_i(x)} \right]  d\mu(x),
\end{equation*}
gives $\F([\Phi])$ the structure of a Gerstenhaber algebra with respect to the ghost degree. Moreover, with 
$$
\tilde \Delta (F)= \sum_{i=1}^N \frac{ \delta_R}{\delta K_{\phi_i}^a(x)}\frac{ \delta_L}{\delta \phi^a_i(x)} (F)
$$
it becomes a BV-algebra. 
\end{prop}
In the physics literature, it is common to write this anti-bracket on the fields generators in terms of the Dirac delta distribution as
$$
(K_{\phi_i}^a(x), \phi^b_{j}(y)) = \delta^{ab} \delta_{ij} \delta(x-y), \quad ( K_{\phi_i}^a(x), K_{\phi_{j}}^a(y)) = 0, \quad (\phi^a_i(x), \phi^b_{j}(y)) = 0
$$
which is then extended to $\F([\Phi])$ using the graded Leibniz property.

\subsection{The comodule BV-algebra of coupling constants and fields}
\label{subsect:comodule}
Since the coupling constants measure the strength of the interactions, we label them by the elements $v \in R_V$ and write accordingly $\lambda_v$. We consider the algebra $A_R$ generated by local functionals in the fields and formal power series (over $\C$) in the coupling constants $\lambda_v$. In other words, we define $A_R :=  \C[[\lambda_{v_1}, \ldots ,\lambda_{v_k}]] \otimes_\C \F([\Phi])$ where $k=|R_V|$. 
The BV-algebra structure on $\F([\Phi])$ defined in the previous section induces a natural BV-algebra structure on $A_R$; we denote the bracket on it by $( \cdot, \cdot)$ as well. 

Recall the notation $H_R$ for the Hopf subalgebra generated by the elements $p_{n_1,\ldots,n_k}(Y_v)$ ($v\in R_V$) and $p_{n_1,\ldots,n_k}(C^\phi)$ ($e\in R_E$) in the Hopf algebra of Feynman graphs.

\begin{thm}
\label{thm:coaction}
The algebra $A_R$ is a comodule BV-algebra for the Hopf algebra $H_R$.
The coaction $\rho: A_R \to A_R \otimes H_R$ is given on the generators by
\begin{align*}
\rho : \lambda_v &\longmapsto \sum_{n_1 \cdots n_k} \lambda_v \lambda_{v_1}^{n_1 } \cdots \lambda_{v_k}^{n_k }  \otimes p_{n_1 \cdots n_k} (Y_v),\\
\rho: \phi & \longmapsto \sum_{n_1 \cdots n_k} \phi~ \lambda_{v_1}^{n_1 } \cdots \lambda_{v_k}^{n_k}  \otimes p_{n_1 \cdots n_k} (C^\phi),
\end{align*}
for $\phi \in \Phi$, while it commutes with partial derivatives on $\phi$. 
\end{thm}
\begin{proof}
Since we work with graded Hopf algebras, it suffices to establish that $ (\rho \otimes 1)\circ \rho= (1 \otimes \Delta) \circ \rho$. We claim that this follows from coassociativity ({i.e.} $(\Delta \otimes 1) \circ \Delta = (1\otimes \Delta)\circ \Delta$) of the coproduct $\Delta$ of $H_R$. Indeed, the first expression very much resembles the form of the coproduct on $Y_v$ as derived in Proposition \ref{prop:cop-Yv}: replacing therein each $Y_{v'}$ on the first leg of the tensor product by $\lambda_{v'}$ and one $\Delta$ by $\rho$ gives the desired result. A similar argument applies to the second expression, using Equation \eqref{cop-Cphi} above. 

Finally, since $C^{K_{\phi_i}} \equiv (C^{\phi_i})^{-1}$ by definition in $H_R$, it follows that $\rho$ respects the bracket and the operator $\tilde\Delta$ and thus the BV-algebra structure. 
\end{proof}

\begin{corl}
The Green's functions $G^v \in H_R$ can be obtained when coacting on the monomial $\int_M \lambda_v \iota(v)(x) d\mu(x)= \int_M\lambda_v \partial_{\vec\mu_1} \phi_{i_1}(x) \cdots \partial_{\vec\mu_M} \phi_{i_M}(x)d \mu(x)$ for some index set $\{ i_1, \ldots, i_M \}$. Explicitly,
\begin{multline*}
\rho\left(\int_M\lambda_v \partial_{\vec\mu_1} \phi_{i_1}(x) \cdots \partial_{\vec\mu_M} \phi_{i_M}(x)\right)
= \!\!\sum_{n_1 \cdots n_k} \!\! \lambda_v \lambda_{v_1}^{n_1} \cdots \lambda_{v_k}^{n_k} \int_M \partial_{\vec\mu_1} \phi_{i_1}(x) \cdots \partial_{\vec\mu_M} \phi_{i_M}(x) \otimes  p_{n_1 \cdots n_k} (G^v).
\end{multline*}
\end{corl}

Combining Theorem \ref{thm:coaction} with Corollary \ref{corl:graph-diffeo} yields an induced coaction on $\C[[\lambda_{v_1}, \ldots, \lambda_{v_k}]]$ of the Hopf algebra dual to the group of diffeomorphisms on $\C^k$ tangent to the identity. The formula for this coaction can be obtained by substituting $a^{(i)}_{n_1 \cdots n_k}$ for $p_{n_1 \cdots n_k} (Y_{v_i})$ in the above formula for $\rho(\lambda_v)$. It induces a group action of $\Diff(\C^k,0)$ on $ \C[[\lambda_{v_1}, \ldots, \lambda_{v_k}]]$ by $f(a) := (1 \otimes f) \rho(a)$ for $f\in \Diff(\C^k,0)$ and $a\in \C[[\lambda_{v_1}, \ldots, \lambda_{v_k}]]$. In fact, we have the following

\begin{prop}
\label{prop:action}
Let $G$ be the group consisting of BV-algebra maps $f: A_R \to A_R$ given on the generators by
\begin{align*}
f ( \lambda_v)&= \sum_{n_1 \cdots n_k} f^v_{n_1 \cdots n_k} \lambda_v \lambda_{v_1}^{n_1 } \cdots \lambda_{v_k}^{n_k }; \qquad (v \in R_V) ,\\
f ( \phi_i)&= \sum_{n_1 \cdots n_k} f^i_{n_1 \cdots n_k} \phi_i \lambda_{v_1}^{n_1 } \cdots \lambda_{v_k}^{n_k }; \qquad (i =1, \ldots, N) , \\
\end{align*}
where $f^v_{n_1 \cdots n_k},f^i_{n_1 \cdots n_k}\in \C$ are such that $f^v_{0 \cdots 0} = f^i_{0\cdots 0} = 1$.
Then the following hold:
\begin{enumerate}
\item The character group $G_R$ of the Hopf algebra $H_R$ generated by $p_{n_1\cdots n_k} (Y_v)$ and $p_{n_1\cdots n_k} (C^\phi)$ with coproduct given in Proposition \ref{prop:cop-Yv}, is a subgroup of $G$.
\item The subgroup $N:= \{ f: f(\lambda_v) =\lambda_v \}$ of $G$ is normal and isomorphic to $(\C[[\lambda_{v_1},\ldots,\lambda_{v_k}]]^\times)^{|R_E|}$.
\item $G \simeq (\C[[\lambda_{v_1},\ldots,\lambda_{v_k}]]^\times)^{|R_E|} \rtimes \Diff(\C^k,0)$.
\end{enumerate}
\end{prop}
\begin{proof}
From Theorem \ref{thm:coaction}, it follows that a character $\chi \in G_R$ acts on $A_R$ as in the above formula upon writing $f^v_{n_1 \cdots n_k} = \chi( p_{n_1\cdots n_k} (Y_v) )$ and $f^i_{n_1 \cdots n_k} = \chi( p_{n_1\cdots n_k} (C^{\phi_i}))$.

For {\it 2.} one checks by explicit computation that $N$ is indeed normal and that each series $f^i$ defines an element in $\C[[\lambda_{v_1}, \ldots, \lambda_{v_k}]]^\times$ of invertible formal power series. 

Then {\it 3.} follows from the existence of a homomorphism from $G$ to $\Diff(\C^k,0)$. It is given by restricting an element $f$ to $\C[[\lambda_{v_1}, \ldots, \lambda_{v_k}]]$. This is clearly the identity map on $\Diff(\C^k,0)$ when considered as a subgroup of $G$ and its kernel is precisely $N$.
\end{proof}

\begin{rem}
Note that the expression for the action of $f \in G$ on the BRST-sources $K_{\phi_i}$ can be derived from the expression for $f(\phi_i)$ above using the fact that $( f(K_{\phi_i})(x) , f(\phi_i)(y) ) = \delta(x-y)$. 
\end{rem}

The action of (the subgroup of) $(\C[[\lambda_{v_1},\ldots,\lambda_{v_k}]]^\times)^{|R_E|} \rtimes \Diff(\C^k,0)$ on $A_R$ has a natural physical interpretation: the invertible formal power series act on every propagating field as wave function renormalization whereas the diffeomorphisms act on the coupling constants $\lambda_1,\ldots,\lambda_k$. The similarity with the semi-direct product structures obtained (via different approaches) in \cite{GKM01} for a scalar field theory and in \cite{BF01,BFK06} for quantum electrodynamics is striking. 
\begin{ex}
Consider again pure Yang--Mills theory with fields $A,\omega,\bar\omega$ and $h$. Then, under the counterterm map $\gamma_-(z) \in G_R$ (cf. Section \ref{sect:birkhoff}) we can identify $(C^A)^2 = G^\glu$ with wave function renormalization for the gluon propagator, and the combination $C^\omega C^{\bar\omega}= G^\gho$ with wave function renormalization for the ghost propagator. The above action of $\gamma_-(z)$ on the fields $A,\omega,\bar\omega$ is thus equivalent to wave function renormalization. 
We will come back to Yang--Mills theories in more detail in Section \ref{sect:ym} below.
\end{ex}

\subsection{The master equation}
\label{sect:master}
The dynamics and interactions in the physical system is described by means of a so-called {\it action} $S$. In our formalism, $S$ will be an element in $A_R$ of polynomial degree $\geq 2$ of the form,
\begin{equation}
\label{eq:action}
S[\phi]= \sum_{e \in R_E}\int d\mu(x)~ \iota(e)(x) + \sum_{v \in R_V} \int d\mu(x)~ \lambda_v ~\iota(v)(x)
\end{equation}
The first sum in $S$ describes the free field theory containing the propagators of the (massless) fields. The second term describes the interactions {\it including} the mass terms. Note that due to the restrictions in the sums, the action has finitely many terms, that is, it is a (local) polynomial functional in the fields rather than a formal power series.

The action $S$ is supposed to be invariant under some group of gauge transformations.\footnote{In addition, it is supposed to be invariant under the symmetry group of the underlying spacetime one works on, typically the Lorentz group. However, these transformations are linear in the fields and will consequently not give rise to non-linear equations such as the master equation discussed here. See for instance \cite{GW96} for more details.} We accomplish this in our setting by imposing the (classical) {\it master equation},
\begin{equation}
\label{eq:master}
( S,S )=0,
\end{equation}
as relations in the BV-algebra $A_R$. 
\begin{prop}
The BV-ideal $I = \langle (S,S) \rangle$ is generated by polynomials in $\lambda_v$ ($v \in R_V$), independent of the fields $\phi \in \Phi$. 
\end{prop}
\begin{proof}
Let us write the master equation for the Lagrangian as a polynomial in $A_R$:
$$
(S,S) = \sum c_{i_1 \cdots i_N} \int_M \partial_{\vec\mu_1} \phi_{i_1}(x) \cdots \partial_{\vec\mu_N} \phi_{i_N} (x) d\mu(x)\in A_R
$$ 
with $c_{i_1 \cdots i_N} \in \C[\lambda]$ a polynomial independent of the fields $\phi$. For $I$ to be a BV-ideal it has to satisfy $(a, I )\subset I$ for any $a \in A_R$. The following property allows us to project onto each individual term in the above polynomial:
\begin{multline*}
\left( \int_M f(x) K_{\phi_i}(x)d\mu(x), \int_M \partial_{\vec\mu_1} \phi_{i_1}(y) \cdots \partial_{\vec\mu_N} \phi_{i_N} (x) d\mu(y) \right)=\\ 
\sum_{\begin{smallmatrix} k \text{ s.t.} \\ i_k = i \end{smallmatrix} } 
(\pm)
\int_M \partial_{\vec\mu_k} f(x) \partial_{\vec\mu_1} \phi_{i_1}(x) \cdots \widehat{\partial_{\vec\mu_k} \phi_{i_k}(x)} \cdots \partial_{\vec\mu_N} \phi_{i_N} (x) d\mu(x).
\end{multline*}
Here $~\widehat{ }~$ means that this factor is absent and $f \in C_c^\infty(M)$ is a test function. Note that $\Loc(E)$ is indeed a $C^\infty_c(M)$-module. Iterating this property, we infer that
\begin{gather*}
\left( \int_M f_1(x_1) K_{\phi_{i_1}}(x_1) d\mu(x_1), \left( \cdots \left( \int_M f_N(x_N) K_{\phi_{i_N}}(x_N) d\mu(x_N), (S,S) \right)\cdots \right)\right) 
\propto c_{i_1 \cdots i_N} F([f]).
\end{gather*}
with $F([f])$ a local functional of the test functions $f_1, \cdots, f_N$. Since these are arbitrary, it follows that $c_{i_1 \cdots i_N} \in I$, in other words, $I$ is already generated by the coefficients of the polynomial $(S,S)$, as claimed.
\end{proof}

We still denote the image of the action $S$ in $A_R/I$ under the quotient map by $S$; it satisfies the master equation \eqref{eq:master} with the brackets as defined before. If we make the natural assumption that $S$ is at most of order one in the BRST-sources, we can write
\begin{equation}
\label{eq:BRSTaction}
S= S_0[\lambda_v, \phi_i] + \sum_{i=1}^N \sum_{a=1}^{\rk E_i} \int d\mu(x) (s \phi_i)^a(x) K^a_{\phi_i}(x).
\end{equation}
with $s\phi_i$ dictated by the previous form of $S$. Of course, this is the familiar BRST-differential acting on the field $\phi_i$ as a graded derivation and obviously satisfies $s \phi_i (x)= (S, \phi_i(x))$. As usual, validity of the master equation $(S,S)=0$ implies that $s$ is nilpotent:
$$
s^2 (\phi_i) = (S, (S, \phi_i)) = \pm ((S,S),\phi_i) = 0
$$
using the graded Jacobi identity. Moreover, the action $S_0$ depending on the fields is BRST-closed, {i.e.} $sS_0 =0$, which follows by considering the part of the master equation $(S,S)=0$ that is independent of the BRST-sources.

\bigskip

The following result establishes an action and coaction on the quotient BV-algebra $A_R/I$. 
\begin{thm}
\label{thm:group-ideal}
Let $G^I_R$ be the (closed) subgroup of $G_R$ defined in Proposition \ref{prop:action} consisting of diffeomorphisms $f$ that leave $I$ invariant, {i.e.} such that $f(I) \subset I$. 
\begin{enumerate}
\item The group $G_R^I$ acts on the quotient BV-algebra $A_R/I$.
\item The ideal in $H_R$ defined by
\begin{equation}
\label{eq:J}
J := \left\{ X \in H_R: f(X) = 0 \text{ for all } f \in G^I_R \right\}
\end{equation}
is a Hopf ideal.
\end{enumerate}
Consequently, $G_R^I\simeq\Hom_\C(H_R/J,\C)$ and the quotient Hopf algebra $\tilde H_R = H_R/J$ coacts on $A_R/I$.
\end{thm}
\begin{proof}
First observe that $G_R^I$ is closed since it can be given as the zero-set of polynomials in $H_R$. Indeed, following \cite[Lemma 12.4]{Wat79} we can write 
$$
\rho(w_j) = \sum_{k \in \mathcal{K}} f(a_{kj}) w_k = \sum_{k \in \mathcal{K} - \mathcal{I}} f(a_{kj}) w_k  + \sum_{i \in \mathcal{I}} f(a_{ij}) w_i
$$
where $\{ w_k : k \in \mathcal{K} \}$ is a (countable) basis for $A_R$ and $\{ w_i : i \in \mathcal{I} \}$ a basis for $I$. Thus, $f$ should satisfy the equations $f(a_{kj}) =0$ for  $k \in \mathcal{K}$ and $ i \in \mathcal{I}$.

For {\it 2.}, we adopt the standard practice in algebraic geometry to relate (closed) subspaces to (radical) ideals. In the present case, we have a one-to-one correspondence between closed subspaces of $\Hom_\C(H_R,\C)$ and radical ideals in the algebra $H_R$ as follows: to each subspace $G$ one associates a ideal $J_G$ (which is prime and hence radical) by the above formula \eqref{eq:J} and vice versa, for every such ideal $J$ there is a subspace $G_J := 
\Hom_\C(H/J, \C)$. By \cite[Proposition 1.2]{Har77} it follows that $G_{J_G} = G$ and $J_{G_J}=J$. Furthermore, if $G$ carries a group structure (as is the case for $G_R^I$), the algebra $H/J_G$ is in fact a Hopf algebra which implies that $J_G$ is a Hopf ideal. 
\end{proof}
We denote the coaction of $\tilde H_R:=H_R/J$ on $A_R$ by $\tilde \rho$; it is given explicitly by
\begin{equation}
\label{coaction-proj}
\tilde \rho(a+I) = \left(\pi_I \otimes \pi_J \right) \rho(a),
\end{equation}
for $a \in A_R$; also, $\pi_I$ and $\pi_J$ are the projections onto the quotient algebra and Hopf algebra by $I$ and $J$ respectively.

Let us now justify the origin of the explicit Hopf ideals that we have encountered in the previous section in the case that all coupling constants coincide. This happens for instance in the case of Yang--Mills theory with a simple gauge group, which is discussed in Section \ref{sect:ym}. 
In general, we make the following definition.
\begin{defn} 
A theory defined by $S$ is called {\rm simple} when the following holds modulo the ideal $\langle \lambda_v  \rangle_{\val(v)=2}$:
\begin{equation}
\label{eq:simple-ideal}
I = \langle \lambda_{v'}^{N(v)-2} - \lambda_v^{N(v')-2} \rangle_{\val(v),\val(v')>2} 
\end{equation}
\end{defn}
In other words, if we put the mass terms in $S$ to zero, then the ideal $I$ should be generated by the differences $\lambda_{v'}^{N(v)-2} - \lambda_v^{N(v')-2}$ for vertices with valence greater than 2. We denote by $I'$ the ideal in Eq. \eqref{eq:simple-ideal} modulo $\langle \lambda_v \rangle_{\val(v)=2}$. A convenient choice of generators for $I'$ is the following. Fix a vertex $v \in R_V$ of valence three,\footnote{We suppose that there exists such a vertex; if not, the construction works equally well by choosing the vertex of lowest valence that is present in the set $R_V$.} and define $g:=\lambda_v$ as the `fundamental' coupling constant. Then $I'$ is generated by $\lambda_v $ with $\val(v)=2$ and $\lambda_{v'} - g^{N(v')-2}$ with $\val(v')>2$. Recall the ideal $J'$ from the previous section.

\begin{thm}
\label{thm:subgroup}
Let $S$ define a simple theory in the sense described above. 
\begin{enumerate}
\item The subgroup $G^{I'}$ of diffeomorphisms that leave $I'$ invariant is isomorphic to $\Hom_\C(H_R/J',K)$.
\item The Hopf algebra $H_R/J'$ coacts on $\C[[g,\phi]]:=A_R/I'$ via the map 
\begin{align*}
\tilde\rho': g &\longmapsto  \sum_{l=0}^\infty g^{2l+1}  \otimes q_l(X),\\
\tilde\rho': \phi &\longmapsto \sum_{l=0}^\infty g^{2l} \phi \otimes q_l(C^{\phi}).
\end{align*}
\end{enumerate}
\end{thm}
\begin{proof}
From the proof of Theorem \ref{thm:group-ideal} we see that {\it 1.} is equivalent to showing that $G_R^{I'} \isom (G_R)_{J'}$. Indeed, $(G_R)_{J'}$ is the subgroup of characters on $H_R$ that vanish on $J' \subset H_R$, which is isomorphic to $\Hom_\C(H_R/J',K)$. On the generators of $I'$, an element $f \in G_R$ acts as
\begin{gather*}
f\left( \lambda_{v'} -g^{N(v')-2} \right) =  \sum_{n_1, \ldots, n_k} \lambda_{v_1}^{n_1} \cdots\lambda_{v_k}^{n_k} \bigg[  \lambda_v f\left( p_{n_1,\ldots,n_k}(Y_{v'}) \right) 
- g^{N(v')-2}  f( p_{n_1,\ldots,n_k}(Y_v^{N(v')-2}))\bigg],
\end{gather*}
where $v$ is the chosen vertex of valence 3 corresponding to $g$. We will reduce this expression by replacing $\lambda_{v_i}$ by $g^{N(v_i)-2}$, modulo terms in $I'$. Together with Lemma \ref{lma:rel-degrees} this yields
$$
f\left( \lambda_{v'} -g^{N(v')-2} \right) = \sum_{l=0}^\infty g^{2l + N(v')-2} ~ f \left(q_l\left( Y_{v'} - Y_v^{N(v')-2}\right) \right)  \mod I'
$$
The requirement that this is an element in $I'$ is equivalent to the requirement that $f$ vanishes on $q_l ( Y_{v'} - Y_v^{N(v')-2})$, {i.e.} on the generators of $J'$, establishing the isomorphism $G_R^{I'} \simeq (G_R)_{J'}$.

For {\it 2.}, one can easily compute
$$
\rho(I') \subset I' \otimes H_R + A_R \otimes J'
$$
so that $H_R/J'$ coacts on $A_R$ by projecting onto the two quotient algebras (as in Eq. \eqref{coaction-proj}).
\end{proof}

\begin{corl}
The group $G_R^{I'}$ acts on $A_R/I'$ as a subgroup of $(\C[[g]]^\times)^{|R_E|} \rtimes \Diff(\C,0)$.
\end{corl}
This last result has a very nice physical interpretation: the invertible formal power series act on the $|R_E|$ propagating fields as wave function renormalization whereas the diffeomorphisms act on one fundamental coupling constant $g$. We will appreciate this even more in the next section where we discuss the renormalization group flow.

\subsection{Renormalization group}
We will now establish a connection between the group of diffeographisms and the renormalization group \`a la Gell'Mann and Low \cite{GL54}. This group describes the dependence of the renormalized amplitudes $\phi_+(z)$ on a mass scale that is implicit in the renormalization procedure. In fact, in dimensional regularization, in order to keep the loop integrals $d^{4-z} k$ dimensionless for complex $z$, one introduces a factor of $\mu^z$ in front of them, where $\mu$ has dimension of mass and is called the {\it unit of mass}. For a Feynman graph $\Gamma$, Lemma \ref{lma:rel-degrees} shows that this factor equals $\mu^{z \sum_{v} (N(v)-2)) \delta_v(\Gamma)/2}$ reflecting the fact that the coupling constants appearing in the action get replaced by 
$$
\lambda_v \mapsto \mu^{z \sum_{v} (N(v)-2))/2}\lambda_v
$$
for every vertex $v \in R_V$. 

As before, the Feynman rules define a loop $\gamma_\mu: C \to G\equiv G(\C)$, which now depends on the mass scale $\mu$. Consequently, there is a Birkhoff decomposition for each $\mu$:
$$
\gamma_\mu(z) = \gamma_{\mu,-}(z)^{-1} \gamma_{\mu,+}(z); \qquad (z \in C),
$$
As was shown in \cite{CK00}, the negative part $\gamma_{\mu,-}(z)$ of this Birkhoff decomposition is independent of the mass scale, that is
$$
\frac{\partial}{\partial \mu} \gamma_{\mu,-}(z) = 0. 
$$
Hence, we can drop the index $\mu$ and write $\gamma_{-}(z):=\gamma_{\mu,-}(z)$. In terms of the generator $\theta_t$ for the one-parameter subgroup of $G(K)$ corresponding to the grading $l$ on $H$, we can write 
$$
\gamma_{e^t \mu (z) } = \theta_{tz} \left(\gamma_\mu(z) \right), \qquad (t \in \R).
$$
A proof of this and the following result can be found in \cite{CK00}.

\begin{prop}
The limit 
$$
F_t := \lim_{z \to 0} \gamma_-(z) \theta_{tz} \left( \gamma_-(z)^{-1} \right)
$$
exists and defines a $1$-parameter subgroup of $G$ which depends polynomially on $t$ when evaluated on an element $X \in H$. 
\end{prop}
In physics, this 1-parameter subgroup goes under the name of {\it renormalization group}. In fact, using the Birkhoff decomposition, we can as well write
$$
\gamma_{e^t \mu, +}(0) = F_t ~ \gamma_{\mu,+}(0), \qquad (t \in \R). 
$$
This can be formulated in terms of the generator $\beta := \frac{d}{dt} F_t |_{t=0}$ of this 1-parameter group as
\begin{equation}
\label{eq:beta}
\mu \frac{\partial}{\partial \mu} \gamma_{\mu,+}(0) = \beta \gamma_{\mu,+}(0). 
\end{equation}
Let us now establish that this is indeed the beta-function familiar from physics by exploring how it acts on the coupling constants $\lambda_v$. First of all, although the name might suggest otherwise, the coupling constants depend on the energy or mass scale $\mu$. Recall the action of $G_R$ on $\C[[\lambda_{v_1}, \ldots, \lambda_{v_k}]]$ defined in the previous section. In the case of $\gamma_{\mu,+}(0) \in G_R$, we define the (renormalized) {\it coupling constant at scale $\mu$} to be
$$
\lambda_v(\mu) = \gamma_{\mu,+}(0)(\lambda_v). 
$$
This function of $\mu$ (with coefficients in $\C[[\lambda_v]]$) satisfies the following differential equation:
\begin{equation*}
\beta \left( \lambda_v(\mu) \right) = \mu \frac{\partial}{\partial \mu} \left(\lambda_v(\mu) \right)
\end{equation*}
which follows easily from Eq. \eqref{eq:beta}. This is exactly the renormalization group equation expressing the flow of the coupling constants $\lambda_v$ as a function of the energy scale $\mu$. 
Moreover, if we extend $\beta$ by linearity to the action $S$ of Eq. \eqref{eq:action}, we obtain Wilson's continuous renormalization equation \cite{Wil75}:
$$
\beta(S(\mu)) = \mu \frac{\partial}{\partial \mu} \left( S(\mu) \right)
$$
This equation has been explored in the context of renormalization Hopf algebras in \cite{GKM04,KM08}.

Equation \eqref{eq:beta} expresses $\beta$ completely in terms of $\gamma_{\mu,+}$; as we will now demonstrate, this allows us to derive that in the case of a simple theory all $\beta$-functions coincide. First, recall that the maps $\gamma_{\mu}$ are the Feynman rules dictated by $S$ in the presence of the mass scale $\mu$, which we suppose to satisfy the master equation \eqref{eq:master}. In other words, we are in the quotient of $A_R$ by $I = \langle (S,S) \rangle$. In addition, assume that the theory defined by $S$ is simple. If the regularization procedure respects gauge invariance, it is well-known that the Feynman amplitude satisfy the Slavnov--Taylor identities for the couplings. In terms of the ideal $J'$ defined in the previous section, this means that $\gamma_{\mu} (J')=0$. Since $J'$ is a Hopf ideal (Theorem \ref{thm:hopfideal}), it follows that both $\gamma_{\mu,-}$ and $\gamma_{\mu,+}$ vanish on $J$. Indeed, the character $\gamma$ given by the Feynman rules factorizes through $H_R/J$ for which the Birkhoff decomposition gives two characters $\gamma_+$ and $\gamma_-$ of $H_R/J$. In other words, if the unrenormalized Feynman amplitudes given by $\gamma_\mu$ satisfy the Slavnov--Taylor identities, so do the counterterms and the renormalized Feynman amplitudes. 

In particular, we find with Equation \eqref{eq:beta} that $\beta$ vanishes on the ideal $I'$ in $\C[[\lambda_{v_1}, \ldots, \lambda_{v_k}]]$. This implies the following result, which is well-known in the physics literature:
\begin{prop}
For a simple theory, all $\beta$-functions are expressed in terms of $\beta(g)$ for the fundamental coupling constant $g$:
$$\beta(\lambda_{v}) = \beta(g^{N(v)-2}).$$
\end{prop}

\section{Example: pure Yang--Mills theory}
\label{sect:ym}

Let us now exemplify the above construction in the case of a pure Yang--Mills theory. Let $G$ be a simple Lie group with Lie algebra $\g$. The {\it gauge field} $A$ is a $\g$-valued one-form, that is, a section of $\Lambda^1 \otimes (M \times \g)$. As before, we have in components $A=A_i^a dx^i T^a$ where the $\{ T^a \}$ form a basis for $\g$. The structure constants $\{ f^{ab}_c \}$ of $\g$ are defined by $[T^a, T^b]=f^{ab}_c T^c$ and the normalization is such that $\tr(T^a T^b) = \delta^{ab}$. 

In addition to the gauge fields, there are {\it ghost fields} $\omega$, $\bar\omega$ which are sections of $M \times \g[-1]$ and $M\times \g[1]$, respectively, and we write $\omega = \omega^a T^a$ and $\bar\omega = \bar\omega^a T^a$. The auxiliary field -- also known as the Nakanishi--Lantrup field -- is denoted by $h = h^a T^a$ and is a section of $M \times \g$. 

The form degree and ghost degree of the fields are combined in the {\it total degree} and summarized in the following table:
\begin{center}
\begin{tabular}{|l|r|r|r|r|}
\hline
& $A$ & $\omega$ & $\bar\omega$ & $h$\\ \hline
ghost degree 	&0    &$+1$ &$-1$ &0	\\ \hline
form degree 	&$+1$ &$0$  &$0$  &0	\\ \hline
total degree	&$+1$ &$+1$ &$-1$ &0	\\ \hline
\end{tabular}
\end{center}

We introduce BRST-sources for each of the above fields, $K_A, K_\omega, K_{\bar\omega}$ and $K_h$. The shift in ghost degree is illustrated by the following table:
\begin{center}
\begin{tabular}{|l|r|r|r|r|}
\hline
& $K_A$ & $K_\omega$ & $K_{\bar\omega}$ & $K_h$\\ \hline
ghost degree 	&$-1$    &$-2$ &$0$ &$-1$	\\ \hline
form degree 	&$+1$ &$0$  &$0$  &0	\\ \hline
total degree	&$0$ &$-2$ &$0$ &$-1$	\\ \hline
\end{tabular}
\end{center}
With these degrees, we can generate the algebra of local forms $\Loc(\Phi)$, which decomposes as before into $\Loc^{(p,q)}(\Phi)$ with $p$ the form degree and $q$ the ghost degree. The total degree is then $p+q$ and $\Loc(\Phi)$ is a graded Lie algebra by setting 
$$
[ X, Y ] = X Y - (-1)^{\deg(X)\deg(Y)} Y X,
$$
with the grading given by this total degree. This bracket should not be confused with the anti-bracket defined on local functionals in Section \ref{sect:BV}. The present graded Lie bracket is of degree $0$ with respect to the total degree, that is, $\deg([X,Y]) = \deg(X) + \deg(Y)$. It satisfies graded skew-symmetry, the graded Leibniz identity and the graded Jacobi identity:
\begin{align*}
&[X,Y] = - (-1)^{\deg(X)\deg(Y)} [Y,X], \\ 
& [XY,Z] = X [ Y,Z] + (-1)^{\deg(Y)\deg(Z)} [X,Z]Y.\\
&(-1)^{\deg(X)\deg(Z)} [ [ X,Y],Z] + \text{(cyclic perm.)} = 0
\end{align*}

\subsection{The Yang--Mills action}

In the setting of Section \ref{sect:master}, the action $S$ for pure Yang--Mills theory is the local functional 
\begin{gather}
S =
\int_M \tr \bigg[
- dA * dA - \lambda_{A^3} dA * [A,A] - \frac{1}{4} \lambda_{A^4} [A,A] * [A,A] 
- A* dh + d \bar\omega * d \omega
+ \frac{1}{2} \xi h * h 
\\ \nonumber 
\quad + \lambda_{\bar\omega A \omega} d \bar\omega * [A,\omega]
- \left( \langle d \omega ,K_A   \rangle 
+\lambda_{A \omega K_A} \langle [A,\omega] , K_A\rangle  
+ \langle h, K_{\bar\omega} \rangle 
+ \frac{1}{2}  \lambda_{\omega^2 K_\omega} \langle [\omega,\omega], K_\omega \rangle \right) *1 \bigg]
\end{gather}
where $*$ denotes the Hodge star operator and $\xi$ is the so-called gauge fixing (real) parameter. Also $\langle \cdot,\cdot \rangle$ denotes the pairing between 1-forms and vector fields (or 0-forms and 0-forms).
In contrast with the usual formula for the action in the literature, we have inserted the different coupling constants $\lambda_v$ for each of the interaction monomials in the action. We will now show that validity of the master equation $(S,S)=0$ implies that all these coupling constants are expressed in terms of one single coupling. 

First, using Eq. \eqref{eq:BRSTaction} we derive from the above expression the BRST-differential on the generators
\begin{align*}
s A = - d \omega - \lambda_{A \omega K_A} [A,\omega],\qquad
s \omega = -\frac{1}{2} \lambda_{\omega^2 K_\omega} [\omega,\omega],\qquad
s \bar\omega = - h,\qquad
s h =0 
\end{align*}
The BRST-differential is extended to all of $\Loc^{(p,q)}(\Phi)$ by the graded Leibniz rule, and imposing it to anti-commute with the exterior derivative $d$. Actually, rather than on $\Loc(\Phi)$, the BRST-differential is defined on the algebra $\C[[\lambda_{A^3}, \lambda_{A^4} ,\lambda_{\bar\omega A \omega}, \lambda_{A \omega K_A}, \lambda_{\omega^2 K_\omega} ]] \otimes \Loc(\Phi)$. However, in order not to loose ourselves in notational complexities, we denote this tensor product by $\Loc(\Phi)$ as well. 

Now, validity of the master equation implies that $s^2=0$. One computes using the graded Jacobi identity that
$$
s^2(A) = \left( \lambda_{A \omega K_A} - \lambda_{\omega^2 K_\omega} \right) [d\omega,\omega] + \frac{1}{2} \left( \lambda_{A \omega K_A}^2 - \lambda_{A \omega K_A} \lambda_{\omega^2 K_\omega} \right) [A, [\omega,\omega]].
$$
from which it follows that $\lambda_{A \omega K_A} = \lambda_{\omega^2 K_\omega}$. 
Thus, with this relation the $s$ becomes a differential, and actually forms -- together with the exterior derivative -- a bicomplex in which $s \circ d + d \circ s = 0$. 
\begin{figure}[t!]
$$
\xymatrix{ 
&\vdots  & \vdots & \vdots &\\
&\Loc^{(0,1)} \ar[u]_s \ar[r]_d  &\Loc^{(1,1)} \ar[u]_s \ar[r]_d  &\Loc^{(2,1)}\ar[u]_s \ar[r]_d & \cdots\\
&\Loc^{(0,0)} \ar[u]_s \ar[r]_d  &\Loc^{(1,0)} \ar[u]_s \ar[r]_d  &\Loc^{(2,0)}\ar[u]_s \ar[r]_d & \cdots\\
&\Loc^{(0,-1)} \ar[u]_s \ar[r]_d  &\Loc^{(1,-1)} \ar[u]_s \ar[r]_d  &\Loc^{(2,-1)}\ar[u]_s \ar[r]_d & \cdots\\
& \vdots \ar[u]_s   & \vdots \ar[u]_s  & \vdots \ar[u]_s & 
}
$$
\end{figure}

\bigskip

Next, the master equation implies that $s S_0 =0$ and a lengthy computation yields for the first three terms in $S_0$ that
\begin{gather*}
 s \left(- dA * dA - \lambda_{A^3} dA * [A,A] - \frac{1}{4} \lambda_{A^4} [A,A] * [A,A]\right) = \hspace{2cm}\\
\hspace{2cm} 2 \left(\lambda_{A \omega K_A} - \lambda_{A^3} \right) dA * [A,d\omega] 
 + (\lambda_{A^4} - \lambda_{A^3} \lambda_{A \omega K_A} ) [d\omega,A] *[A,A] \\
\hspace{2cm} + \lambda_{A \omega K_A} \left( - dA * dA - \lambda_{A^3} dA * [A,A] - \frac{1}{4} \lambda_{A^4} [A,A] * [A,A]~,~ \omega\right).
\end{gather*}
The last term is a commutator on which the trace vanishes and one is thus left with the equalities $\lambda_{A \omega K_A} = \lambda_{A^3}$ and $\lambda_{A^4} = \lambda_{A^3} \lambda_{A \omega K_A}$. The remaining terms in $S_0$ yield under the action of $s$
\begin{multline*}
s \left(
- A* dh + d \bar\omega * d \omega
+ \frac{1}{2} \xi h * h 
+ \lambda_{\bar\omega A \omega} d \bar\omega * [A,\omega]
\right)= \\( \lambda_{A \omega K_A} - \lambda_{\bar\omega A \omega} ) [A,\omega]* d h 
+ (\lambda_{\omega^2 K_\omega} - \lambda_{\bar\omega A \omega}) d \bar\omega * [d\omega,\omega].
\end{multline*}
Thus, the master equation implies $\lambda_{A \omega K_A} = \lambda_{\bar\omega A \omega}$ and $\lambda_{\omega^2 K_\omega} = \lambda_{\bar\omega A \omega}$. 

Finally, if we write $g = \lambda_{A^3}$, the master equation implies that 
\begin{equation}
\label{ym-couplings}
\lambda_{A^4} = g^2 \text{ and } \lambda_{\bar\omega A \omega} = \lambda_{A \omega K_A} = \lambda_{\omega^2 K_\omega}=g.
\end{equation}
This motivates our definition of a simple theory in Section \ref{sect:master} above. Imposing these relations reduces the action $S$ to the usual
\begin{gather*}
S = \int_M \tr \bigg[ - F * F - A * dh + d \bar\omega * d\omega + g d \bar\omega * [A,\omega] + \frac{1}{2} \xi h * h 
+ s A * K_A + s \omega * K_\omega + s \bar\omega * K_{\bar\omega}
\bigg]
\end{gather*}
with the field strength $F$ given by $F= dA + \tfrac{g}{2} [A,A]$ and the BRST-differential now given by 
$$
s A = -d \omega - g [A,\omega], \qquad s \omega = -\frac{1}{2} g [\omega,\omega], \qquad s \bar\omega = -h, \qquad s h =0.
$$
The extension to include fermions is straightforward, leading to similar expressions of the corresponding coupling constants in terms of $g$.

\subsection{The action of $G_R$}
As alluded to before, when the counterterm map -- seen as an element in $G_R$ -- acts on the action $S$, it coincides with wave function renormalization. Let us make this precise in the present case. Clearly, wave function renormalization is given by the following factors:
\begin{gather*}
Z_A = \gamma_-(z)(G^\glu); \quad Z_\omega = Z_{\bar\omega} =  \gamma_-(z)(G^\gho).
\end{gather*}
With this definition and Theorem \ref{thm:coaction} we find that $\gamma_-(z)$ acts as
\begin{align*}
\gamma_-(z) \cdot (dA * dA)  &= \gamma_-(z) \left( (C^A)^2 \right) dA * dA = Z_A ~ dA * dA \\ 
\gamma_-(z) \cdot (d\bar\omega * d\omega)  &= \gamma_-(z) (C^\omega C^{\bar\omega} ) d\bar\omega * d\omega= Z_\omega ~ d\bar\omega * d\omega
\end{align*}
by definition of the $C^\phi$'s. This is precisely wave function renormalization for the gluon and ghost fields. Thus, renormalizing through the coefficients $\gamma_-(z)(C^\phi)$ -- although more appropriate for the BV-formalism -- is completely equivalent to the usual wave function renormalization (see also \cite[Section 6]{Ans94}).

By construction, the terms $-A * dh$ and $\langle h, K_{\bar\omega} \rangle$ do not receive radiative corrections. Indeed, this follows from the relations:
$$
C^b C^A = 1; \qquad C^{K_{\bar\omega}} C^b = 1,
$$
in $H_R$. Consequently, $G_R$ -- and in particular the counterterm map $\gamma_-(z)$ -- acts as the identity on these monomials. 

In fact, one realizes that $S_0= \gamma_-(z)\cdot S$ is the {\it renormalized action}, and since $\gamma_-(z) \in G_R$ acts as a BV-algebra map, also $S_0$ satisfies the master equation $(S_0,S_0)=0$. This will be further explored in future work.

\subsection{The Slavnov--Taylor identities}
We now use Theorem \ref{thm:subgroup} to obtain the relations between the Green's function in Yang--Mills equations that are induced by the above master equation $(S,S)=0$. In fact, the action $S$ defines a simple theory in the sense defined before and Equation \eqref{ym-couplings} implies that the following relations hold in the quotient Hopf algebra $H_R/J'$:
\begin{align*}
Y_\gluq = (Y_\gluc)^2 \text{ and } Y_\gluc = Y_\ghoglu  = Y_\gluBRST = Y_\ghoBRST.
\end{align*}
In terms of the Green's functions the most relevant read
\begin{align*}
\frac{G^\gluq}{(G^\glu)^2} = \left( \frac{(G^\gluc)}{(G^\glu)^{3/2}} \right)^2, \quad \frac{G^\gluc}{(G^\glu)^{3/2}} = \frac{G^\ghoglu}{(G^\glu)^{1/2} G^\gho}, \text{ and } G^\ghoglu = G^\gluBRST.
\end{align*}
These are precisely the Slavnov--Taylor identities for the coupling constants for pure Yang--Mills theory with a simple Lie group.

\section{Outlook}
\label{section:outlook}
The connection we have established between renormalization Hopf algebras for gauge theories and the BV-algebras generated by the relevant fields and coupling constants paves the way for an incorporation of the full BV-formalism in the context of Hopf algebras. This formalism is very powerful in that it can handle theories that are renormalizable `in the modern sense'. Instead of restricting to Lagrangians with a finite number of terms, one allows here a formal series admitting an infinite number of counterterms; the only condition is then the (quantum) master equation. We expect that in this case the group $(\C[[g]]^\times)^{|R_E|} \rtimes \Diff(\C,0)$ encountered above gets replaced by the semi-direct product of so-called {\it canonical transformations} with the diffeomorphism group. Here canonical transformations are automorphisms of the BV-algebra $A_R$, thus respecting the bracket.

Another perspective of our work is in the direction of BRST-quantization. A description of the BRST-formalism -- typically exploited in the physical literature involving functional methods -- in the Hopf algebraic setting would elucidate the role it plays in renormalization of gauge theories.

There are potential applications of the current setup in the approach taken by Hollands in \cite{Hol07} to perturbatively quantizing Yang--Mills theories on curved spacetimes. There, Ward identities are formulated in terms of functionals as well and renormalization is supposed to respect them. Motivated by the present construction in momentum space, it is expected that these identities induce Hopf ideals in the Hopf algebra of \cite{Pin00} describing Epstein--Glaser renormalization. 
Another subject we have not touched is gauge theories with spontaneous symmetry breaking. It would be interesting to study renormalization of such theories in the present setup.

Finally, the necessity of the Slavnov--Taylor-like identities in the work of Kreimer on quantum gravity \cite{Kre07,Kre08} is quite intriguing. In fact, Theorem \ref{thm:hopfideal} can be extended \cite{KS09} to the so-called core Hopf algebra that was introduced in \cite{BK08}, consisting of graphs with vertices of any valence.

We postpone the study of the effective action in the Hopf algebraic setting to our next paper. The Zinn--Justin equation it satisfies will play a similar role as the (classical) master equation \eqref{eq:master} in imposing identities between the 1PI Green's functions, albeit now for {\it any} interaction and not only for those represented by the set $R_V$ as discussed in Sect. \ref{sect:master}.  Also, we will connect with the usual order-by-order in the loop number approach to renormalization of gauge theories that is taken in the physics literature.

\section*{Acknowledgements}
The author would like to thank Caterina Consani, George Elliott, Johan Martens and Jim Stasheff for their kind invitations in October 2007, where much of this work was initiated. I want to thank Glenn Barnich, Detlev Buchholz, Alain Connes, Klaus Fredenhagen, Eugene Ha, Dirk Kreimer, Matilde Marcolli and Jack Morava for valuable discussions and remarks. Finally, the Hausdorff Research Institute for Mathematics in Bonn is acknowledged for their hospitality during the final stages of this work. 



\begin{thebibliography}{10}

\bibitem{Ans94}
D.~Anselmi.
\newblock {Removal of divergences with the Batalin-Vilkovisky formalism}.
\newblock {\em Class. Quant. Grav.} 11 (1994)  2181--2204.

\bibitem{BKUY08}
G.~van Baalen, D.~Kreimer, D.~Uminsky, and K.~Yeats.
\newblock {The QED beta-function from global solutions to Dyson-Schwinger
  equations}, arXiv:0805.0826.

\bibitem{BBH94}
G.~Barnich, F.~Brandt, and M.~Henneaux.
\newblock Local {BRST} cohomology in the antifield formalism. {I}. {G}eneral
  theorems.
\newblock {\em Commun. Math. Phys.} 174 (1995)  57--92.

\bibitem{BBH95}
G.~Barnich, F.~Brandt, and M.~Henneaux.
\newblock Local {BRST} cohomology in the antifield formalism. {II}.
  {A}pplication to {Y}ang-{M}ills theory.
\newblock {\em Commun. Math. Phys.} 174 (1995)  93--116.

\bibitem{BV81}
I.~A. Batalin and G.~A. Vilkovisky.
\newblock {Gauge Algebra and Quantization}.
\newblock {\em Phys. Lett.} B102 (1981)  27--31.

\bibitem{BV83}
I.~A. Batalin and G.~A. Vilkovisky.
\newblock Feynman rules for reducible gauge theories.
\newblock {\em Phys. Lett.} B120 (1983)  166--170.

\bibitem{BV84}
I.~A. Batalin and G.~A. Vilkovisky.
\newblock {Quantization of Gauge Theories with Linearly Dependent Generators}.
\newblock {\em Phys. Rev.} D28 (1983)  2567--2582.

\bibitem{BRS74}
C.~Becchi, A.~Rouet, and R.~Stora.
\newblock The abelian {H}iggs-{K}ibble model. {U}nitarity of the {S} operator.
\newblock {\em Phys. Lett.} B52 (1974)  344.

\bibitem{BRS75}
C.~Becchi, A.~Rouet, and R.~Stora.
\newblock Renormalization of the abelian {H}iggs-{K}ibble model.
\newblock {\em Commun. Math. Phys.} 42 (1975)  127--162.

\bibitem{BRS76}
C.~Becchi, A.~Rouet, and R.~Stora.
\newblock Renormalization of gauge theories.
\newblock {\em Annals Phys.} 98 (1976)  287--321.

\bibitem{BK04}
C.~Bergbauer and D.~Kreimer.
\newblock {The Hopf algebra of rooted trees in Epstein-Glaser renormalization}.
\newblock {\em Annales Henri Poincare} 6 (2005)  343--367.

\bibitem{BK05}
C.~Bergbauer and D.~Kreimer.
\newblock Hopf algebras in renormalization theory: {L}ocality and
  {D}yson-{S}chwinger equations from {H}ochschild cohomology.
\newblock {\em IRMA Lect. Math. Theor. Phys.} 10 (2006)  133--164.

\bibitem{BK08}
S.~Bloch and D.~Kreimer.
\newblock {Mixed Hodge Structures and Renormalization in Physics}.
\newblock {\em Commun. Num. Theor. Phys. 2.} 4 (2008)  637--718.

\bibitem{BF01}
C.~Brouder and A.~Frabetti.
\newblock Renormalization of {QED} with planar binary trees.
\newblock {\em Eur. Phys. J.} C19 (2001)  715--741.

\bibitem{BFK06}
C.~Brouder, A.~Frabetti, and C.~Krattenthaler.
\newblock Non-commutative {H}opf algebra of formal diffeomorphisms.
\newblock {\em Adv. Math.} 200 (2006)  479--524.

\bibitem{BruF00}
R.~Brunetti and K.~Fredenhagen.
\newblock {Microlocal analysis and interacting quantum field theories:
  Renormalization on physical backgrounds}.
\newblock {\em Commun. Math. Phys.} 208 (2000)  623--661.

\bibitem{CK99}
A.~Connes and D.~Kreimer.
\newblock Renormalization in quantum field theory and the {R}iemann- {H}ilbert
  problem. {I}: {T}he {H}opf algebra structure of graphs and the main theorem.
\newblock {\em Comm. Math. Phys.} 210 (2000)  249--273.

\bibitem{CK00}
A.~Connes and D.~Kreimer.
\newblock Renormalization in quantum field theory and the {R}iemann- {H}ilbert
  problem. {II}: The beta-function, diffeomorphisms and the renormalization
  group.
\newblock {\em Commun. Math. Phys.} 216 (2001)  215--241.

\bibitem{DF00}
M.~D{\"u}tsch and K.~Fredenhagen.
\newblock Perturbative algebraic field theory, and deformation quantization.
\newblock In {\em Mathematical physics in mathematics and physics ({S}iena,
  2000)}, volume~30 of {\em Fields Inst. Commun.}, pages 151--160. Amer. Math.
  Soc., Providence, RI, 2001.

\bibitem{FGV05}
H.~Figueroa, J.~M. Gracia-Bondia, and J.~C. Varilly.
\newblock {Fa\`a di Bruno Hopf algebras}, math/0508337.

\bibitem{Foi07}
L.~Foissy.
\newblock Fa\`a di {B}runo subalgebras of the {H}opf algebra of planar trees
  from combinatorial {D}yson-{S}chwinger equations.
\newblock {\em Adv. Math.} 218 (2008)  136--162.

\bibitem{GL54}
M.~Gell-Mann and F.~E. Low.
\newblock Quantum electrodynamics at small distances.
\newblock {\em Phys. Rev.} 95 (1954)  1300--1312.

\bibitem{Ger63}
M.~Gerstenhaber.
\newblock The cohomology structure of an associative ring.
\newblock {\em Ann. of Math.} 78 (1963)  267--288.

\bibitem{GKM01}
F.~Girelli, T.~Krajewski, and P.~Martinetti.
\newblock {Wave-function renormalization and the {H}opf algebra of {C}onnes and
  {K}reimer}.
\newblock {\em Mod. Phys. Lett.} A16 (2001)  299--303.

\bibitem{GKM04}
F.~Girelli, T.~Krajewski, and P.~Martinetti.
\newblock {An algebraic {B}irkhoff decomposition for the continuous
  renormalization group}.
\newblock {\em J. Math. Phys.} 45 (2004)  4679--4697.

\bibitem{GW96}
J.~Gomis and S.~Weinberg.
\newblock Are nonrenormalizable gauge theories renormalizable?
\newblock {\em Nucl. Phys.} B469 (1996)  473--487.

\bibitem{Har77}
R.~Hartshorne.
\newblock {\em Algebraic Geometry}.
\newblock Number~52 in Graduate Texts in Mathematics. Springer-Verlag, New
  York, 1977.

\bibitem{Hol07}
S.~Hollands.
\newblock Renormalized quantum {Y}ang-{M}ills fields in curved spacetime.
\newblock {\em Rev. Math. Phys.} 20 (2008)  1033--1172.

\bibitem{HW02}
S.~Hollands and R.~M. Wald.
\newblock {On the renormalization group in curved spacetime}.
\newblock {\em Commun. Math. Phys.} 237 (2003)  123--160.

\bibitem{KM08}
T.~Krajewski and P.~Martinetti.
\newblock Wilsonian renormalization, differential equations and {H}opf
  algebras.
\newblock arXiv:0806.4309.

\bibitem{Kre98}
D.~Kreimer.
\newblock On the {H}opf algebra structure of perturbative quantum field
  theories.
\newblock {\em Adv. Theor. Math. Phys.} 2 (1998)  303--334.

\bibitem{Kre05}
D.~Kreimer.
\newblock Anatomy of a gauge theory.
\newblock {\em Ann. Phys.} 321 (2006)  2757--2781.

\bibitem{Kre07}
D.~Kreimer.
\newblock {A remark on quantum gravity}.
\newblock {\em Ann. Phys.} 323 (2008)  49--60.

\bibitem{Kre08}
D.~Kreimer.
\newblock {Not so non-renormalizable gravity}, arXiv:0805.4545.

\bibitem{KS09}
D.~Kreimer and W.~D. van Suijlekom.
\newblock Work in progress.

\bibitem{KY06}
D.~Kreimer and K.~Yeats.
\newblock An \'etude in non-linear {D}yson-{S}chwinger equations.
\newblock {\em Nuclear Phys. B Proc. Suppl.} 160 (2006)  116--121.

\bibitem{Pin00}
G.~Pinter.
\newblock {The Hopf algebra structure of Connes and Kreimer in Epstein-Glaser
  renormalization}.
\newblock {\em Lett. Math. Phys.} 54 (2000)  227--233.

\bibitem{Sau89}
D.~J. Saunders.
\newblock {\em The Geometry of Jet Bundles}.
\newblock Cambridge University Press, 1989.

\bibitem{Sta97}
J.~Stasheff.
\newblock Deformation theory and the {B}atalin-{V}ilkovisky master equation.
\newblock In {\em Deformation theory and symplectic geometry (Ascona, 1996)},
  volume~20 of {\em Math. Phys. Stud.}, pages 271--284. Kluwer Acad. Publ.,
  Dordrecht, 1997.

\bibitem{Sui07}
W.~D. van Suijlekom.
\newblock Renormalization of gauge fields: {A} {H}opf algebra approach.
\newblock {\em Commun. Math. Phys.} 276 (2007)  773--798.

\bibitem{Sui07b}
W.~D. van Suijlekom.
\newblock Multiplicative renormalization and {H}opf algebras.
\newblock In O.~Ceyhan, Y.-I. Manin, and M.~Marcolli, editors, {\em Arithmetic
  and geometry around quantization}. Birkh\"auser Verlag, Basel, 2008.
\newblock [arXiv:0707.0555].

\bibitem{Sui07c}
W.~D. van Suijlekom.
\newblock Renormalization of gauge fields using {H}opf algebras.
\newblock In J.~T. B.~Fauser and E.~Zeidler, editors, {\em Quantum Field
  Theory}. Birkh\"auser Verlag, Basel, 2008.
\newblock [arXiv:0801.3170].

\bibitem{Tyu75}
I.~V. Tyutin.
\newblock Gauge invariance in field theory and statistical physics in operator
  formalism.
\newblock LEBEDEV-75-39.

\bibitem{Wat79}
W.~C. Waterhouse.
\newblock {\em Introduction to Affine Group Schemes}.
\newblock Springer, New York, 1979.

\bibitem{Wil75}
K.~G. Wilson.
\newblock Renormalization group methods.
\newblock {\em Advances in Math.} 16 (1975)  170--186.

\end{thebibliography}
\newcommand{\noopsort}[1]{}

\end{fmffile}
\end{document}